\documentclass[prl,twocolumn,superscriptaddress]{revtex4-2}
\usepackage{graphicx,amssymb,amsmath,bbold,bm,xcolor,hyperref,mathtools}
\usepackage[normalem]{ulem}
\graphicspath{{img/}}
\usepackage[caption=false]{subfig}
\usepackage{usebib}
\bibinput{main}

\def\arucl{$\alpha$-RuCl$_3$}

\begin{document}
	
\title{Extended Quantum Spin Liquid with Spinon-like Excitations in an Anisotropic Kitaev-Gamma Model}
\author{Matthias Gohlke}
\affiliation{Theory of Quantum Matter Unit, Okinawa Institute of Science and Technology Graduate University,
  Onna-son, Okinawa 904-0495, Japan}
\author{Jose Carlos Pelayo}
\affiliation{Quantum Systems Unit, Okinawa Institute of Science and Technology Graduate University,
  Onna-son, Okinawa 904-0495, Japan}
\author{Takafumi Suzuki}
\affiliation{Graduate School of Engineering, University of Hyogo, Himeji 671-2280, Japan}
\date{\today}     
	
\begin{abstract}
    The characterization of quantum spin liquid phases in Kitaev materials has been a subject of intensive studies over the recent years,
    both theoretically and experimentally.
    Most theoretical studies have focused on an isotropically interacting model with its coupling strength being equivalent on each bond in an attempt to simplify the problem.
    Here, we study an extended spin-1/2 Kitaev-$\Gamma$ model on a honeycomb lattice with an additional tuning parameter
    that controls the coupling strength on one of the bonds:
    we connect the limit of isolated Kitaev-$\Gamma$ chains, which is known to exhibit an emergent $SU(2)_1$ Tomonaga-Luttinger liquid phase [Yang et al. Phys. Rev. Lett. {\bf 124}, 147205 (2020)],
    to the two-dimensional model.
    We report on an instance, in which the Tomonaga-Luttinger liquid persists for finite inter-chain coupling.
    A quantum spin liquid phase develops in analogy to \emph{sliding Luttinger liquids}
    that differs from the Kitaev spin liquid.
    This quantum spin liquid phase features spinon-like excitations similar to those of the antiferromatnetic Heisenberg chain.
    We use numerical Exact Diagonalization and Density Matrix Renormalization Group on various cluster geometries in a complementary way to overcome finite-size limitations.

\end{abstract}
	
\maketitle


Quantum spin liquids (QSL) have become an important research subject in condensed matter physics due to their exotic emergent properties \cite{balents_spin-liquids_2010,savary_quantum_2017,knolle_field_guide_to_QSL_2019}.
Following the 'More is different' philosophy by Anderson \cite{anderson_rvb_1987}, competing interactions---or frustration--- can give rise to novel emergent features:
fractional excitations, topological order, emergent gauge fields, anyonic exchange statistics, etc.
Kitaev's honeycomb model \cite{kitaev_anyons_2006} is a paradigmatic spin-$1/2$ model in this context,
due to being exactly solvable and featuring a QSL ground state
in terms of itinerant Majorana fermions in a static $\mathbb Z_2$ gauge field.
The bond-dependent Kitaev spin-exchange is realized in certain 
magnets with strong spin-orbit coupling \cite{jackeli_mott_2009}.
This mechanism, however, introduces additional spin exchanges \cite{jackeli_mott_2009,JefferyPRL2014} that spoil the exact solvability of the Kitaev model. 
Many candidate Kitaev materials have been proposed 
\cite{trebst_2017,winter_2017,hermanns_physics_of_Kitaev_2018,motome_2020}
among which \arucl~\cite{plumb_aRuCl3_2014} has gained much attention due to a putative QSL phase in an in-plane magnetic field 
\cite{baek_field-induced_QSL_2017,banerjee_excitations_field-induced_QSL_2018}, and even more so, 
since the measurement of a half-quantized thermal hall effect was reported suggesting the existence of emergent Majorana fermions 
\cite{kasahara_majorana_2018,yamashita_sample_dependence_2020,yokoi_half-integer_2021,bruin_robustness_2022}. 

The spin-1/2 Kitaev-$\Gamma$ (K$\Gamma$) model on the honeycomb lattice with ferromagnetic Kitaev exchange
and positive symmetric off-diagonal $\Gamma$ exchange has been proposed as a minimal model for \arucl \cite{Kejing2017_PRL,Janssen2017_PRB, wang_PRB_2017,Winter2017_NatMat}.
Many different methods have been applied, yet no clear understanding of its quantum ground state has emerged. 
Among the suggested ground state phases are not only magnetically ordered states, such as zigzag \cite{wang_JKG_VMC_2019}, ferromagnet \cite{wang_JKG_VMC_2019,lee_magneticfield_2019, buessen_KGamma_FRG_2021}, six-sublattice \cite{lee_magneticfield_2019},
or incommensurate spiral order \cite{JefferyPRL2014,wang_JKG_VMC_2019,buessen_KGamma_FRG_2021},
but also quantum paramagnetic phases such as a putative gapped QSL \cite{GohlkePRB2018}, 
a lattice-nematic paramagnet \cite{lee_magneticfield_2019,gohlke_lattice-nematic_2020},
or a gapless QSL with multiple Majorana-Dirac nodes \cite{wang_JKG_VMC_2019}.
 
We like to change the perspective and focus instead on an aspect rarely included in theoretical works on Kitaev materials:
we additionally tune the strength of the spin exchange spatially, ranging from a limit of uncoupled chains to spatially equal, yet still strongly anisotropic, spin exchange.
In fact, such spatial anisotropy may either be intrinsic due to a reduced symmetry of the underlying lattice, such as $C2/m$ \cite{johnson_monoclinic_2015,cao_low-T_structure_2016, janssen_magnon_dispersion_2020, vilmos_magnetoelastic_coupling_2022} instead of a full $C_3$ rotational symmetry, 
or spatial anisotropy can be induced by applying external pressure or strain \cite{kaib_magnetoelastic_coupling_2021}, which possibly realizes various different QSL \cite{wang_anisoKG_VMC_2020}. 

The one-dimensional limit features an emergent Tomonaga-Luttinger liquid (TLL)
with the same critical properties as the antiferromagnetic Heisenberg (AFH) chain \cite{yang_KGchain_2020}.
Since the TLL is a critical state, most TLLs are unstable under adding small inter-chain coupling \cite{giamarchi_book,schulz_dynamics_of_1996}.
An exception is \emph{sliding Luttinger liquids} \cite{emery_smectic_metal_state_2000,vishwanath_nonfermiliquid_2001,mukhopadhyay_sliding_luttinger_2001} 
that occur if the inter-chain coupling competes with the dominant correlations along the chains.
As a consequence, long-range order is suppressed.

Here, we argue that a similar mechanism arises in the strongly anisotropic K$\Gamma$ model:
The TLL phase of the K$\Gamma$ chain turns into an extended QSL phase.
This QSL retains the dominant algebraic correlations along the chains
and features spinon-like excitations that are characteristic for the AFH chain. 
This contrasts the emergent Majorana fermions and $Z_2$ fluxes of the KSL.

\begin{figure*}[tb]
        \centering
        \includegraphics[width=\linewidth]{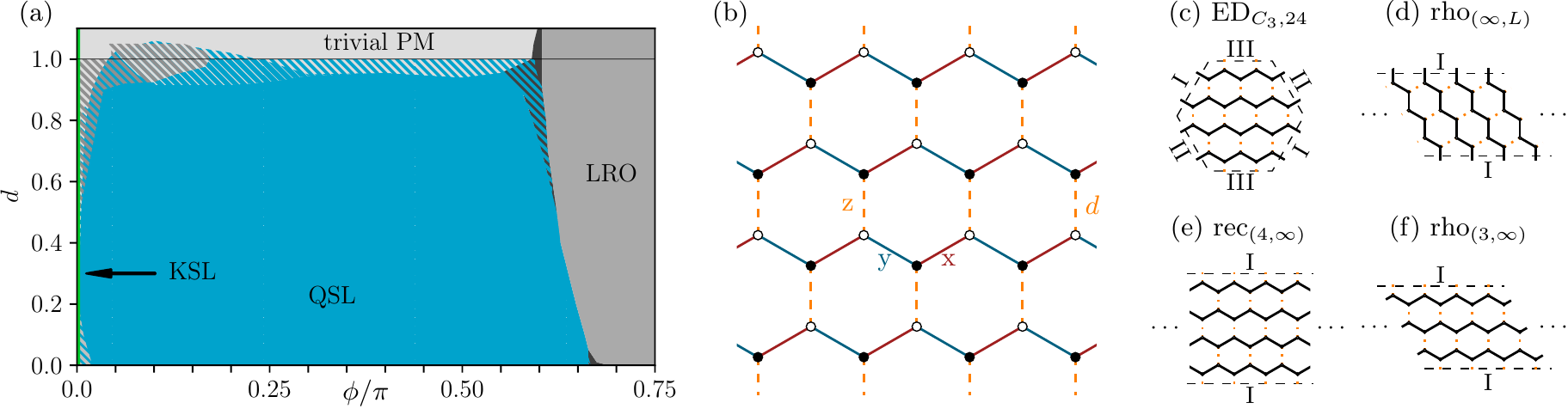}
        \caption{\label{fig:Phasediagram}
            (a) Schematic ground state phase diagram of the anisotropic K$\Gamma$ model.
            The QSL phase is highlighted in blue,
            while the KSL exist only in a tiny range next to $\phi=0$.
            LRO refers to a magnetically ordered phase with non-coplanar $90^\circ$ order.
            Hatched areas illustrate where results of DMRG and ED differ.
            (b) Illustration of the K$\Gamma$ model on the honeycomb lattice [see Eq.~\ref{eqn:KGHam}] with bond-dependent interactions
            and additional modification of the coupling along $z$ by a factor $d$.  
            We use various geometries: (c) $C_3$-symmetric 24-site cluster within ED,
            (d) rhombic unit cell with finite K$\Gamma$-chains ($L=\{6,12,18\}$) coupled along the infinite direction,
            (e) rectangular unit cell with four infinite K$\Gamma$ chains,
            and (f) rhombic unit cell with three infinite K$\Gamma$-chains.
            Geometries (d-f) are employed within DMRG.}
\end{figure*}

Remarkably, we find the QSL phase with TLL character to extend within a wide range of $\Gamma>0$  with ferromagnetic Kitaev exchange,
and up to the $C_3$-symmetric K$\Gamma$-model with spatially equal spin-exchange strength ($C_3$ limit), 
cf.~Fig.~\ref{fig:Phasediagram}(a). 
We base our reasoning on the following key observations upon tuning the inter-chain coupling up to the $C_3$ limit:
(i) absence of a phase transition in a wide range of $K/\Gamma$,
(ii) absence of a spectral gap,
(iii) strong similarities of the dynamical spin-structure factor (DSF) with the AFH chain,
and (iv) suppression of the effective inter-chain coupling. 

\textit{Anisotropic K$\Gamma$ Model.}
We consider the spin-1/2 K$\Gamma$ model on the honeycomb lattice with $x$, $y$, and $z$ bonds [see Fig.~\ref{fig:Phasediagram}(b)]. Its spin exchange is described by the Hamiltonian 
\begin{align}
    \mathcal{H} =& -K \sum_{\mathclap{\langle i,j\rangle_{\gamma=x,y}}} S^\gamma_i S^\gamma_j + \Gamma \sum_{\mathclap{\substack{\langle i,j\rangle_{\gamma=x,y} \\ \alpha,\beta \neq \gamma}}} \left[ S^\alpha_i S^\beta_j + S^\beta_i S^\alpha_j \right] \nonumber \\
                & -dK \sum_{\mathclap{\langle i,j\rangle_{\gamma=z}}} S^\gamma_i S^\gamma_j + d\Gamma \sum_{\mathclap{\substack{\langle i,j\rangle_{\gamma=z} \\ \alpha,\beta \neq \gamma}}} \left[ S^\alpha_i S^\beta_j + S^\beta_i S^\alpha_j \right], \label{eqn:KGHam}
\end{align}
where $d$ determines the strength of the K$\Gamma$-exhange on the $z$ bond;
$d=1$ refers to equal exchange strength along each bond restoring the $C_3$ lattice-rotation symmetry,
while $d=0$ refers to the chain limit.
Here, we are interested in the range $0 \le d \lessapprox 1$ connecting both limits.
Furthermore, we introduce a trigonometric parameterization $K=\cos \phi$ and $\Gamma=\sin \phi$
and focus on the range $0 \le \phi/\pi \le 3/4$. 

%

Yang et al. \cite{yang_KGchain_2020, yang_JKGchain_2020} have extensively studied the K$\Gamma$-chain limit and identified two relevant sublattice transformations:
a six-sublattice transformation mapping the K$\Gamma$-chain to either the ferromagnetic Heisenberg (FMH) chain at $\phi/\pi=3/4$ or the AFH chain at $\phi/\pi=-1/4$,
and a three-sublattice transformation mapping $\Gamma \mapsto -\Gamma$
such that the phase diagram is symmetric about the pure Kitaev limits, $\phi \mapsto -\phi$.
An emergent $SU(2)_1$ TLL phase has been reported in a wide range, $0 < \phi/\pi < \pm 0.66$ \cite{yang_KGchain_2020}, around the dual AFH points, $\phi/\pi=\pm1/4$.
While the $SU(2)$ symmetry is a property of the Hamiltonian at the dual points, 
the system retains the $SU(2)$ symmetry as an emergent property at any other $\phi$ within the TLL phase.

\textit{Methods.}
We make use of different numerical techniques and their respective cluster geometries [cf.~Fig.~\ref{fig:Phasediagram}~(c-f)] in a complementary way.
Exact diagonalizations (ED) enables us to study the K$\Gamma$ model on a $C_3$-symmetric cluster with 24 sites. 
In the chain limit, the 24-site cluster decomposes into two chains with 12 sites each implying large finite-size gaps.
With Density Matrix Renormalization Group (DMRG) \cite{white_dmrg_1992} and the infinite matrix product state (iMPS) \cite{mcculloch_infinite_2008,phien_infinite_2012} framework, on the other hand, we can treat chains of infinite length eliminating their finite-size gap.
When comparing with ED, it is instructive to also consider a geometry with finite chains in DMRG.
In the remainder of the text, we will refer to each geometry by the label given in Fig.~\ref{fig:Phasediagram}~(c-f).
On general grounds, we expect ED to be more reliable in obtaining the positions of the phase transitions near the $C_3$ limit, $d\approx 1$, due to maintaining the lattice symmetries. 
On the other hand, DMRG and iMPS are not limited by the finite-size gaps of K$\Gamma$ chains at finite length and are expected to be more reliable for weakly coupled chains, $d \ll 1$. 
Furthermore, we compute dynamcial spin-spin correlations employing 
a time evolution of matrix product states \cite{zaletel_tmpo_2015,gohlke_dynamics_JK_2017} for the $\text{rho}_{(3,\infty)}$ geometry with three coupled chains.

\textit{Ground State.}
\begin{figure*}[tb]
    \begin{center}
        \includegraphics[width=\hsize]{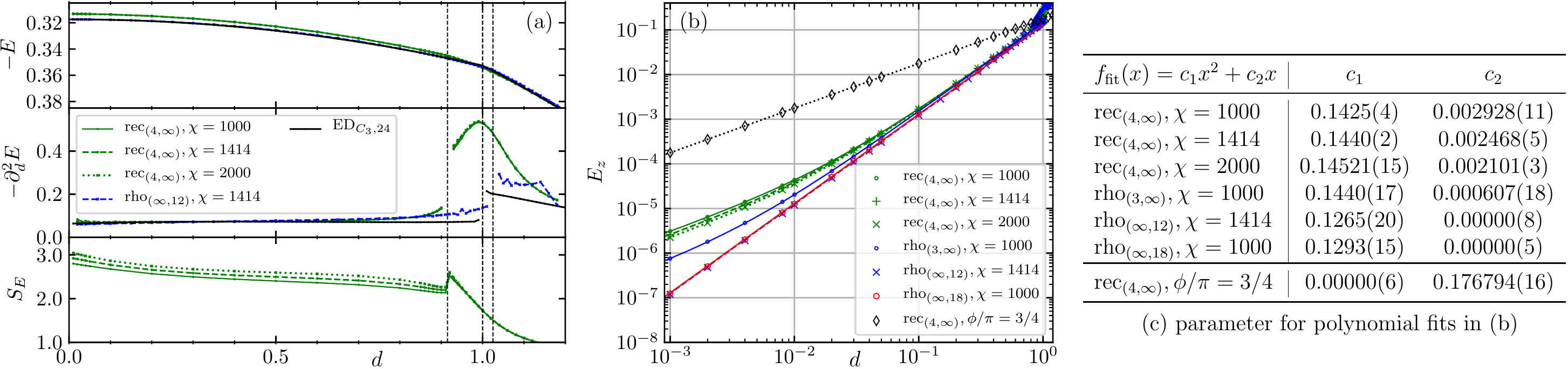}
        \caption{\label{fig:phasediagram_cut} 
            (a) Representative cut through the phase diagram, Fig.~\ref{fig:Phasediagram}, with constant $\phi/\pi=1/4$ and varying inter-chain coupling $d$.
            ED (black) features a single transition at $d^{ED}_{c}=1$ close to $d^{12}_{c}=1.025$ of  $\text{rec}_{(\infty,12)}$, whereas $\text{rec}_{(4,\infty)}$ exhibits two transitions at $d^\infty_{c,1} \approx 0.92$ and $d^\infty_{c,2} \approx 0.99$.
            For $d<d^\infty_{c,2}$, $\text{rec}_{(4,\infty)}$ (green) features a $\chi$-dependence of the entanglement entropy of a bipartition, $S_E$, suggesting a gapless phase.  
            (b) Contribution to the ground state energy from the z-bond, $E_z$, scaling as $E_z \propto d^2$,
            while $E_z$ scales linearly, $E_z \propto d$, in the vertex phase near $\phi/\pi = 3/4$.
            Lines in (b) are fits with parameters listed in table (c).
        }
    \end{center}
\end{figure*}
We start by focusing on the ground state properties at fixed $\phi/\pi=1/4$, cf.~Fig.~\ref{fig:phasediagram_cut}.
Near the chain-limit, $\text{ED}_{C_3,24}$ obtains a lower ground state energy in agreement with the finite-chain geometry $\text{rho}_{(\infty,12)}$ in DMRG.
The finite-size gap is sizable for both geometries.
Shortly before approaching the $C_3$ limit, $\text{rec}_{(4,\infty)}$ with infinite chains obtains a lower ground state energy
until eventually the ground state energies for all geometries converge to a single line for $d \gtrapprox 1.1$ 
and finite-size effects become negligible.

The second derivative $\partial_d^2 E$ features only a single phase transition at $d^{ED}_{c}=1$ for $\text{ED}_{C_3,24}$ and at $d^{12}_{c}=1.025$ for $\text{rho}_{(\infty,12)}$.
The results on $\text{rec}_{(4,\infty)}$ exhibit two features:
A kink originating from a first-order transition near $d^\infty_{c,1}=0.92$ and a broad peak around $d^\infty_{c,2} \approx 0.99$. 
The entanglement entropy of a bipartition, $S_E$, is particularly enhanced for $d<d^\infty_{c,1}$, 
which connects to the chain limit.
Here, $S_E$ depends on the bond dimension, $\chi$, in a way that is consistent with a gapless phase;
$S_E(\chi)$ follows equidistant lines if $\chi$ is increased exponentially, $\chi= \{1000,1414,2000\} \approx 2^\frac{n}{2} \cdot 1000$ with $n=0,1,2$.
$S_E$ drops off quickly upon increasing $d$ beyond $d^\infty_{c,1}$ and $S_E$ ceases to depend on $\chi$, which implies a gapped phase. 
A spectral gap is expected for $d > 1$ due to its adiabatic connection to a product state of triplet states \cite{YamadaT2020}
with quadrupolar order \cite{andreev_spin_1984}.
In the case of $\text{rec}_{(4,\infty)}$, finite-size effects are most pronounced for $d\approx 1$,
likely resulting in a finite-size gap already for $d<1$ and the sequence of transitions observed here.

We turn now to small inter-chain coupling, $d \rightarrow 0$. 
Figure~\ref{fig:phasediagram_cut}(b) shows the energy at the inter-chain bond, $E_z$, as a function of $d$.
By fitting the function $f(d) = c_1 d^2 + c_2 d$, we observe that $E_z$ does depend quadratically on $d$ at $\phi/\pi=1/4$ within the QSL phase.
This behaviour is most evident for the finite-chain geometries $\text{rho}_{(\infty,L)}$ with $L=\{12,18\}$,
while the infinite-chain geometries, $\text{rho}_{(3,\infty)}$ and $\text{rec}_{(4,\infty)}$,
have small corrections to the quadratic behaviour which further reduce as $\chi$ is increased.
In contrast, a linear scaling is expected for the magnetically long-range ordered phase at $\phi/\pi=3/4$, as confirmed in Fig.~\ref{fig:phasediagram_cut}(b).

Finding a quadratic, rather than a linear, dependence on $d$ suggests that the dominant correlations along the chains compete with the inter-chain coupling. 
While successive application of the six- and the three-sublattice rotation
maps the K$\Gamma$ chains to AFH chains along the $x$ and $y$ bonds \cite{yang_KGchain_2020},
the exchange on $z$ bonds maps to a Heisenberg-like diagonal exchange with a peculiar, positionally dependent modulation of the signs for each diagonal component.
This modulation is not compatible with the dominant N\'eel-like correlations along the chain \cite{long_paper}. 
Likewise, a perturbative treatment of coupled chains appears to be challenging due to the high symmetry of the K$\Gamma$ chain \cite{yang_coupled_JKGchains_2022}.

\textit{Extent of the QSL phase.}
Based on the ground-state energy $E=E(d,\phi)$ for the Hamiltonian \eqref{eqn:KGHam}, its first derivative $\partial_{d} E$ ($\partial_{\phi} E$), and its second derivative $\partial_{d}^2 E$ ($\partial_{\phi}^2 E$) against $d$ ($\phi$),
we map out the extent of the QSL phase originating from the TLL, cf.~Fig.~\ref{fig:Phasediagram}(a).
The ferromagnetic KSL at $\phi=0$ is very fragile against the $\Gamma$ interaction,
which is consistent with earlier results at $d=1$~\cite{JefferyPRL2014,GohlkePRB2018,ZhangBatistaPRB2021}.
Results for $\text{ED}_{C_3,24}$ suggest a direct transition between KSL and QSL for $0<d<0.9$,
while for $\text{rec}_{(4,\infty)}$ an intermediate phase exists with a dimerization along the $x$ and $y$ bonds for $\phi/\pi<0.02$ at small $d \lessapprox 0.2$,
as well as an intermediate trivial paramagnet for $0.3 \lessapprox d \lessapprox 1$ and $\phi/\pi < 0.05$.

In the chain limit, $d=0$, the TLL phase appears for $0$ ($0.02$) $< \phi/\pi<0.66$ in ED (DMRG). 
A long-range ordered phase with $D_4$ symmetry (D4FM) stabilizes at $0.69 \lessapprox \phi/\pi<0.88$ around the FMH dual point \cite{yang_KGchain_2020}. 
When $d>0$ D4FM turns into a long-range ordered phase with non-coplanar $90^\circ$ order.
In contrast to Refs.~\cite{yang_KGchain_2020,yang_JKGchain_2020}, 
we find an additional region with strong incommensurate correlations between TLL and D4FM,
which survives small inter-chain coupling $d\lessapprox 0.05$.
For $0.05 \lessapprox d < d_{c,\text{IM2}} \approx 0.8$ ($0.6$), $\text{ED}_{C_3,24}$ ($\text{rec}_{(4,\infty)}$) shows a direct first-order transition between QSL and the non-coplanar order.
A second intermediate phase stabilizes for $d > d_{c,\text{IM2}}$, 
which is characterized by the absence of local magnetic moments,
and a critical scaling consistent with a gapless spectrum \cite{long_paper}.

Near the $C_3$ limit, $d \approx 1$, ED and DMRG results show qualitative differences. 
$\text{ED}_{C_3,24}$ features a single transition between the QSL and a trivial paramagnet (PM) at $d=1$ in a wide range $0.2 \lessapprox \phi/\pi \lessapprox 0.6$.
The PM phase is connected adiabatically to the triplet-dimer limit at $d\rightarrow \infty$ \cite{YamadaT2020}.
On the other hand, $\text{rec}_{(4,\infty)}$ features two transitions: a kink in $\partial^2_d E$ at $d\approx 0.9$ to $0.95$, and a broad feature near $d\approx 1$.
Both transitions merge into a single kink at $d\approx 0.95$ for $\phi/\pi \gtrapprox 0.3$.

%
\textit{iMPS $\chi$-Scaling.}
In the following, we utilize the procedure outlined in Refs.~\cite{rams_precise_extrapolation_2018,vanhecke_scaling_hypothesis_2019}
to extract the spectral gap  by using the eigenvalues of an iMPS transfer matrix (TM) which contain full information about the equal-time correlations \cite{zauner_transfer_2015,he_signatures_2017}.
For Hamiltonians with only local interactions, the equal-time correlations are related to the spectral gap \cite{hastings_locality_2004}, $\xi \sim 1/\Delta$.
This statement has been extended by Zauner et al. \cite{zauner_transfer_2015} to include momentum,
such that $\xi(\bm k) \sim 1/\Delta(\bm k)$ in the vicinity of $\bm k$.
Moreover, if the symmetry upon translation along the cylinder 
circumference is not broken, then for each $k_y=2\pi/L_y$ a set of TM-eigenvalues $\lambda_i$ exists
with a longitudinal momentum $k_x = \arg \lambda_i$ corresponding to the momentum at a minimum of the spectrum. 
Here, we are sorting $\lambda_i$ by their momenta $\bm k = (k_x,k_y)$, and only use $\lambda_i$ for $k_x \approx 2\pi/3$ and $k_y=0$, where we obtain the smallest $\epsilon_{\bm k} = \epsilon_{0,\bm k} = 1/\lambda_1$, corresponding to the dominant correlations.

\begin{figure}[tb]
    \begin{center}
        \includegraphics[width=\linewidth]{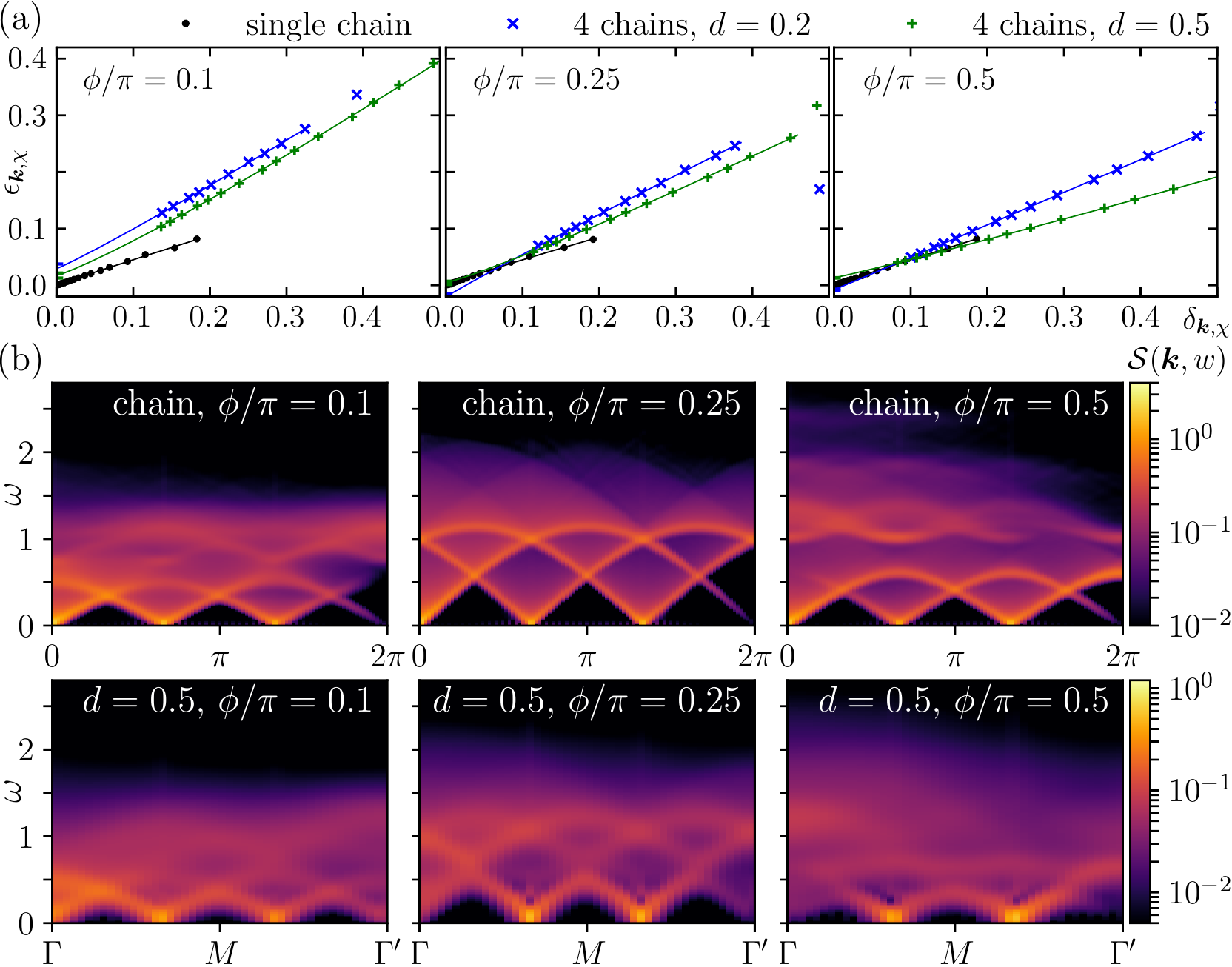}
        \caption{\label{fig:finite-chi_scaling_and_dyn} 
            (a) The spectral gap $\epsilon_k(\chi)$ at $\bm k = (2\pi/3,0)$ extracted from the transfer matrix of the iMPS vanishes linearly with $\delta_k(\chi)$ as $\chi \rightarrow \infty$
            similar to the K$\Gamma$-chain (black).
            We confirm these results at $\phi/\pi = \{0.1,0.25,0.5\}$ using $\text{rec}_{(4,\infty)}$ at $d=0.2$ (blue) and $0.5$ (green).
            (b)  Dynamical spin-structure factor $\mathcal S(\bm k.\omega)$ for the K$\Gamma$-chain (top),
            and $\text{rho}_{(3,\infty)}$ at $d=0.5$ (bottom).
            At $\phi/\pi=1/4$ the K$\Gamma$-chain is equivalent to the AFH chain with its characteristic spinon continuum.
            The six-sublattice transformation results in three instances of the spinon spectrum shifted by $\delta k_x=\pm 2\pi/3$.
            Within the resolution limit ($\sigma_\omega = 0.080$),
            the linear gapless modes and the spinon continuum persists for finite inter chain coupling.
        }
    \end{center}
\end{figure}

We find $\epsilon_k(\chi)$ to decrease as a function of the spacing between the inverse eigenvalues, $\delta_k(\chi) = \sum_i c_i \epsilon_{i,k}$, where $\sum_i c_i = 0$.
The results are shown in Fig.~\ref{fig:finite-chi_scaling_and_dyn}(a).
Solid lines represent fits to the function $f(\delta)=c_1 \delta^{c_2} + c_0$. 
In all cases, we observe an almost linear behaviour with $c_2=1.0(1)$ and $c_0=0.00(3)$ consistent with a gapless nature of the QSL.
For comparison, we include similar data for a single K$\Gamma$ chain whose ground state is a gapless $SU(2)_1$ TLL with a well defined scaling form.
Moreover, we confirm that local magnetic moments vanish in the $\chi \rightarrow \infty$ limit,
ruling out long-range magnetic order \cite{long_paper}. 

\textit{Dynamical properties.}
At $\phi/\pi=1/4$, where the K$\Gamma$ chain is equivalent to the AFH chain,
the excitations are spin-1/2 spinons with the characteristic continuum in the dynamical spin structure factor (DSF) bounded by a double arc and linear dispersion at the gapless points \cite{bethe_1931,yamada_fermi-liquid_1969,mueller_quantum_spin_dynamics_1981,karbach_two-spinon_1997}.
As shown in Fig.~\ref{fig:finite-chi_scaling_and_dyn}(b), the K$\Gamma$ chain features three copies of the AFH chain spectrum due to a tripling of the unit cell by the six-sublattice transformation.
For coupled infinite chains---here, we use $\text{rho}_{(3,\infty)}$ at $d=0.5$---
the DSF retains these features %
\footnote{Apart from an intrinsic broadening set by the longest times reached within the numerical simulations.}.
Within bounds set by the resolution, the DSF remains gapless with linear dispersion at $\bm k = \{\Gamma, \frac{2}{3} M, \frac{4}{3} M, \Gamma'\}$. 
The broad continuum is mostly unchanged apart from a shift of some spectral weight to low energies at $\Gamma'$.
Upon tuning towards either $\Gamma \rightarrow 0$ or $K \rightarrow 0$, the lower edge of the spectrum flattens,
while maintaining the same arc-like lower boundary with linear gapless modes. 
At $\phi/\pi=0.1$, the upper edge of the continuum moves towards lower energies ($\omega_\text{max} \approx 2$) approaching the upper cutoff known for the KSL \cite{knolle_dynamics_2014,knolle_dynamics_2015}.
At $\phi/\pi=0.5$, on the other hand, the upper edge moves to higher energies,
and a second dispersive feature develops at $1 < \omega < 1.4$. 

We interpret the similarities in the DSF between coupled chains and the K$\Gamma$-chain as a signature of the excitations being of the same spinon nature.
Although changing $\phi$ changes the DSF notably, the qualitative behaviour---and, thus, the nature of the excitations---at low-energies remains similar.

\textit{Discussion.}
We have provided numerical evidence for the existence of an extended QSL phase born out of the emergent TLL phase in the one-dimensional limit.
Competing spin exchanges due to the inter-chain coupling lead to frustration, suppress long-range magnetic order,
and stabilize a QSL with spinon-like excitations.
Remarkably, this phase extends up to the $C_3$ limit.

One may argue that additional exchanges, such as Heisenberg, $\Gamma'$, or further-neighbor exchanges,
will act as a singular perturbation and stabilize long-range magnetic order \cite{yang_coupled_JKGchains_2022}.
However, we want to highlight the possible relation of the QSL reported here
with the lattice-nematic paramagnet (NP) found in a K$\Gamma\Gamma'$ model in applied magnetic field along the $[111]$ axis \cite{lee_magneticfield_2019,gohlke_lattice-nematic_2020}.
There, upon adding a small negative $\Gamma'$, the NP phase gives way to zigzag magnetic order at low field,
while NP reappears once the applied magnetic field suppresses the magnetic order.

This has important implications on possible realisations in materials:
(i) the QSL exists in a wide region of mostly ferromagnetic Kitaev and antiferro-like $\Gamma$ exchange and in an applied external magnetic field. When the crystal symmetry remains (close to) $C_3$, applying a magnetic field along the $[111]$ axis (or $c^*$ axis) and tuning the tilting angle appropriately can select the NP state that corresponds to this QSL \cite{gohlke_lattice-nematic_2020}.
(ii) the QSL may be stabilized in Kitaev materials with a from $C_3$ down to $C_2$ reduced space-group symmetry.
Such a reduction can be intrinsic due to a reduced crystal symmetry. Alternatively, applying a uniaxial strain or an external magnetic field will alter the spin-exchange via magnetoelastic coupling \cite{kaib_magnetoelastic_coupling_2021, vilmos_magnetoelastic_coupling_2022}.

\textit{Acknowledgements.}
This work was supported by JSPS KAKENHI (Grants No. 21K03390 and 22K14008) from MEXT, Japan. 
We acknowledge the use of computational resources of the supercomputer Fugaku provided by the RIKEN AICS through the HPCI System Research Project (Project ID: hp210321), of the ISSP Supercomputer Center at the University of Tokyo,
and of the Scientific Computing section of the Research Support Division at the Okinawa Institute of Science and Technology Graduate University (OIST).
M.G.~acknowledges support by the Theory of Quantum Matter Unit at OIST.


\bibliography{main}

\begin{thebibliography}{62}%
\makeatletter
\providecommand \@ifxundefined [1]{%
 \@ifx{#1\undefined}
}%
\providecommand \@ifnum [1]{%
 \ifnum #1\expandafter \@firstoftwo
 \else \expandafter \@secondoftwo
 \fi
}%
\providecommand \@ifx [1]{%
 \ifx #1\expandafter \@firstoftwo
 \else \expandafter \@secondoftwo
 \fi
}%
\providecommand \natexlab [1]{#1}%
\providecommand \enquote  [1]{``#1''}%
\providecommand \bibnamefont  [1]{#1}%
\providecommand \bibfnamefont [1]{#1}%
\providecommand \citenamefont [1]{#1}%
\providecommand \href@noop [0]{\@secondoftwo}%
\providecommand \href [0]{\begingroup \@sanitize@url \@href}%
\providecommand \@href[1]{\@@startlink{#1}\@@href}%
\providecommand \@@href[1]{\endgroup#1\@@endlink}%
\providecommand \@sanitize@url [0]{\catcode `\\12\catcode `\$12\catcode
  `\&12\catcode `\#12\catcode `\^12\catcode `\_12\catcode `\%12\relax}%
\providecommand \@@startlink[1]{}%
\providecommand \@@endlink[0]{}%
\providecommand \url  [0]{\begingroup\@sanitize@url \@url }%
\providecommand \@url [1]{\endgroup\@href {#1}{\urlprefix }}%
\providecommand \urlprefix  [0]{URL }%
\providecommand \Eprint [0]{\href }%
\providecommand \doibase [0]{https://doi.org/}%
\providecommand \selectlanguage [0]{\@gobble}%
\providecommand \bibinfo  [0]{\@secondoftwo}%
\providecommand \bibfield  [0]{\@secondoftwo}%
\providecommand \translation [1]{[#1]}%
\providecommand \BibitemOpen [0]{}%
\providecommand \bibitemStop [0]{}%
\providecommand \bibitemNoStop [0]{.\EOS\space}%
\providecommand \EOS [0]{\spacefactor3000\relax}%
\providecommand \BibitemShut  [1]{\csname bibitem#1\endcsname}%
\let\auto@bib@innerbib\@empty
\bibitem [{\citenamefont {Balents}(2010)}]{balents_spin-liquids_2010}%
  \BibitemOpen
  \bibfield  {author} {\bibinfo {author} {\bibfnamefont {L.}~\bibnamefont
  {Balents}},\ }\bibfield  {title} {\bibinfo {title} {{Spin liquids in
  frustrated magnets}},\ }\href {https://doi.org/10.1038/nature08917}
  {\bibfield  {journal} {\bibinfo  {journal} {Nature}\ }\textbf {\bibinfo
  {volume} {464}},\ \bibinfo {pages} {199} (\bibinfo {year}
  {2010})}\BibitemShut {NoStop}%
\bibitem [{\citenamefont {Savary}\ and\ \citenamefont
  {Balents}(2017)}]{savary_quantum_2017}%
  \BibitemOpen
  \bibfield  {author} {\bibinfo {author} {\bibfnamefont {L.}~\bibnamefont
  {Savary}}\ and\ \bibinfo {author} {\bibfnamefont {L.}~\bibnamefont
  {Balents}},\ }\bibfield  {title} {\bibinfo {title} {{Quantum Spin Liquids: A
  Review}},\ }\href {https://doi.org/10.1088/0034-4885/80/1/016502} {\bibfield
  {journal} {\bibinfo  {journal} {Rep. Prog. Phys.}\ }\textbf {\bibinfo
  {volume} {80}},\ \bibinfo {pages} {016502} (\bibinfo {year}
  {2017})}\BibitemShut {NoStop}%
\bibitem [{\citenamefont {Knolle}\ and\ \citenamefont
  {Moessner}(2019)}]{knolle_field_guide_to_QSL_2019}%
  \BibitemOpen
  \bibfield  {author} {\bibinfo {author} {\bibfnamefont {J.}~\bibnamefont
  {Knolle}}\ and\ \bibinfo {author} {\bibfnamefont {R.}~\bibnamefont
  {Moessner}},\ }\bibfield  {title} {\bibinfo {title} {{A Field Guide to Spin
  Liquids}},\ }\href {https://doi.org/10.1146/annurev-conmatphys-031218-013401}
  {\bibfield  {journal} {\bibinfo  {journal} {Annual Review of Condensed Matter
  Physics}\ }\textbf {\bibinfo {volume} {10}},\ \bibinfo {pages} {451}
  (\bibinfo {year} {2019})}\BibitemShut {NoStop}%
\bibitem [{\citenamefont {{Anderson}}(1987)}]{anderson_rvb_1987}%
  \BibitemOpen
  \bibfield  {author} {\bibinfo {author} {\bibfnamefont {P.~W.}\ \bibnamefont
  {{Anderson}}},\ }\bibfield  {title} {\bibinfo {title} {{The resonating
  valence bond state in La2CuO4 and superconductivity}},\ }\href
  {https://doi.org/10.1126/science.235.4793.1196} {\bibfield  {journal}
  {\bibinfo  {journal} {Science}\ }\textbf {\bibinfo {volume} {235}},\ \bibinfo
  {pages} {1196} (\bibinfo {year} {1987})}\BibitemShut {NoStop}%
\bibitem [{\citenamefont {Kitaev}(2006)}]{kitaev_anyons_2006}%
  \BibitemOpen
  \bibfield  {author} {\bibinfo {author} {\bibfnamefont {A.}~\bibnamefont
  {Kitaev}},\ }\bibfield  {title} {\bibinfo {title} {{Anyons in an Exactly
  Solved Model and Beyond}},\ }\href
  {https://doi.org/10.1016/j.aop.2005.10.005} {\bibfield  {journal} {\bibinfo
  {journal} {Ann. Phys. (NY)}\ }\textbf {\bibinfo {volume} {321}},\ \bibinfo
  {pages} {2} (\bibinfo {year} {2006})}\BibitemShut {NoStop}%
\bibitem [{\citenamefont {Jackeli}\ and\ \citenamefont
  {Khaliullin}(2009)}]{jackeli_mott_2009}%
  \BibitemOpen
  \bibfield  {author} {\bibinfo {author} {\bibfnamefont {G.}~\bibnamefont
  {Jackeli}}\ and\ \bibinfo {author} {\bibfnamefont {G.}~\bibnamefont
  {Khaliullin}},\ }\bibfield  {title} {\bibinfo {title} {Mott {{Insulators}} in
  the {{Strong Spin}}-{{Orbit Coupling Limit}}: {{From Heisenberg}} to a
  {{Quantum Compass}} and {{Kitaev Models}}},\ }\href
  {https://doi.org/10.1103/PhysRevLett.102.017205} {\bibfield  {journal}
  {\bibinfo  {journal} {Phys. Rev. Lett.}\ }\textbf {\bibinfo {volume} {102}},\
  \bibinfo {pages} {017205} (\bibinfo {year} {2009})}\BibitemShut {NoStop}%
\bibitem [{\citenamefont {Rau}\ \emph {et~al.}(2014)\citenamefont {Rau},
  \citenamefont {Lee},\ and\ \citenamefont {Kee}}]{JefferyPRL2014}%
  \BibitemOpen
  \bibfield  {author} {\bibinfo {author} {\bibfnamefont {J.~G.}\ \bibnamefont
  {Rau}}, \bibinfo {author} {\bibfnamefont {E.~K.-H.}\ \bibnamefont {Lee}},\
  and\ \bibinfo {author} {\bibfnamefont {H.-Y.}\ \bibnamefont {Kee}},\
  }\bibfield  {title} {\bibinfo {title} {{Generic Spin Model for the Honeycomb
  Iridates beyond the Kitaev Limit}},\ }\href
  {https://doi.org/10.1103/PhysRevLett.112.077204} {\bibfield  {journal}
  {\bibinfo  {journal} {Phys. Rev. Lett.}\ }\textbf {\bibinfo {volume} {112}},\
  \bibinfo {pages} {077204} (\bibinfo {year} {2014})}\BibitemShut {NoStop}%
\bibitem [{\citenamefont {{Trebst}}(2017)}]{trebst_2017}%
  \BibitemOpen
  \bibfield  {author} {\bibinfo {author} {\bibfnamefont {S.}~\bibnamefont
  {{Trebst}}},\ }\bibfield  {title} {\bibinfo {title} {{Kitaev Materials}},\
  }\href@noop {} {\bibfield  {journal} {\bibinfo  {journal} {ArXiv e-prints}\ }
  (\bibinfo {year} {2017})},\ \Eprint {https://arxiv.org/abs/1701.07056}
  {arXiv:1701.07056 [cond-mat.str-el]} \BibitemShut {NoStop}%
\bibitem [{\citenamefont {Winter}\ \emph
  {et~al.}(2017{\natexlab{a}})\citenamefont {Winter}, \citenamefont {Tsirlin},
  \citenamefont {Daghofer}, \citenamefont {van~den Brink}, \citenamefont
  {Singh}, \citenamefont {Gegenwart},\ and\ \citenamefont
  {Valentí}}]{winter_2017}%
  \BibitemOpen
  \bibfield  {author} {\bibinfo {author} {\bibfnamefont {S.~M.}\ \bibnamefont
  {Winter}}, \bibinfo {author} {\bibfnamefont {A.~A.}\ \bibnamefont {Tsirlin}},
  \bibinfo {author} {\bibfnamefont {M.}~\bibnamefont {Daghofer}}, \bibinfo
  {author} {\bibfnamefont {J.}~\bibnamefont {van~den Brink}}, \bibinfo {author}
  {\bibfnamefont {Y.}~\bibnamefont {Singh}}, \bibinfo {author} {\bibfnamefont
  {P.}~\bibnamefont {Gegenwart}},\ and\ \bibinfo {author} {\bibfnamefont
  {R.}~\bibnamefont {Valentí}},\ }\bibfield  {title} {\bibinfo {title}
  {{Models and materials for generalized Kitaev magnetism}},\ }\href
  {https://doi.org/10.1088/1361-648X/aa8cf5} {\bibfield  {journal} {\bibinfo
  {journal} {Journal of Physics: Condensed Matter}\ }\textbf {\bibinfo {volume}
  {29}},\ \bibinfo {pages} {493002} (\bibinfo {year}
  {2017}{\natexlab{a}})}\BibitemShut {NoStop}%
\bibitem [{\citenamefont {Hermanns}\ \emph {et~al.}(2018)\citenamefont
  {Hermanns}, \citenamefont {Kimchi},\ and\ \citenamefont
  {Knolle}}]{hermanns_physics_of_Kitaev_2018}%
  \BibitemOpen
  \bibfield  {author} {\bibinfo {author} {\bibfnamefont {M.}~\bibnamefont
  {Hermanns}}, \bibinfo {author} {\bibfnamefont {I.}~\bibnamefont {Kimchi}},\
  and\ \bibinfo {author} {\bibfnamefont {J.}~\bibnamefont {Knolle}},\
  }\bibfield  {title} {\bibinfo {title} {{Physics of the Kitaev Model:
  Fractionalization, Dynamic Correlations, and Material Connections}},\ }\href
  {https://doi.org/10.1146/annurev-conmatphys-033117-053934} {\bibfield
  {journal} {\bibinfo  {journal} {Annual Review of Condensed Matter Physics}\
  }\textbf {\bibinfo {volume} {9}},\ \bibinfo {pages} {17} (\bibinfo {year}
  {2018})}\BibitemShut {NoStop}%
\bibitem [{\citenamefont {Motome}\ \emph {et~al.}(2020)\citenamefont {Motome},
  \citenamefont {Sano}, \citenamefont {Jang}, \citenamefont {Sugita},\ and\
  \citenamefont {Kato}}]{motome_2020}%
  \BibitemOpen
  \bibfield  {author} {\bibinfo {author} {\bibfnamefont {Y.}~\bibnamefont
  {Motome}}, \bibinfo {author} {\bibfnamefont {R.}~\bibnamefont {Sano}},
  \bibinfo {author} {\bibfnamefont {S.}~\bibnamefont {Jang}}, \bibinfo {author}
  {\bibfnamefont {Y.}~\bibnamefont {Sugita}},\ and\ \bibinfo {author}
  {\bibfnamefont {Y.}~\bibnamefont {Kato}},\ }\bibfield  {title} {\bibinfo
  {title} {{Materials design of Kitaev spin liquids beyond the
  Jackeli–Khaliullin mechanism}},\ }\href
  {https://doi.org/10.1088/1361-648X/ab8525} {\bibfield  {journal} {\bibinfo
  {journal} {J. Phys.: Condens. Matter}\ }\textbf {\bibinfo {volume} {32}},\
  \bibinfo {pages} {404001} (\bibinfo {year} {2020})}\BibitemShut {NoStop}%
\bibitem [{\citenamefont {Plumb}\ \emph {et~al.}(2014)\citenamefont {Plumb},
  \citenamefont {Clancy}, \citenamefont {Sandilands}, \citenamefont {Shankar},
  \citenamefont {Hu}, \citenamefont {Burch}, \citenamefont {Kee},\ and\
  \citenamefont {Kim}}]{plumb_aRuCl3_2014}%
  \BibitemOpen
  \bibfield  {author} {\bibinfo {author} {\bibfnamefont {K.~W.}\ \bibnamefont
  {Plumb}}, \bibinfo {author} {\bibfnamefont {J.~P.}\ \bibnamefont {Clancy}},
  \bibinfo {author} {\bibfnamefont {L.~J.}\ \bibnamefont {Sandilands}},
  \bibinfo {author} {\bibfnamefont {V.~V.}\ \bibnamefont {Shankar}}, \bibinfo
  {author} {\bibfnamefont {Y.~F.}\ \bibnamefont {Hu}}, \bibinfo {author}
  {\bibfnamefont {K.~S.}\ \bibnamefont {Burch}}, \bibinfo {author}
  {\bibfnamefont {H.-Y.}\ \bibnamefont {Kee}},\ and\ \bibinfo {author}
  {\bibfnamefont {Y.-J.}\ \bibnamefont {Kim}},\ }\bibfield  {title} {\bibinfo
  {title} {{$\ensuremath{\alpha}\ensuremath{-}{\mathrm{RuCl}}_{3}$: A
  spin-orbit assisted Mott insulator on a honeycomb lattice}},\ }\href
  {https://doi.org/10.1103/PhysRevB.90.041112} {\bibfield  {journal} {\bibinfo
  {journal} {Phys. Rev. B}\ }\textbf {\bibinfo {volume} {90}},\ \bibinfo
  {pages} {041112(R)} (\bibinfo {year} {2014})}\BibitemShut {NoStop}%
\bibitem [{\citenamefont {Baek}\ \emph {et~al.}(2017)\citenamefont {Baek},
  \citenamefont {Do}, \citenamefont {Choi}, \citenamefont {Kwon}, \citenamefont
  {Wolter}, \citenamefont {Nishimoto}, \citenamefont {van~den Brink},\ and\
  \citenamefont {B\"uchner}}]{baek_field-induced_QSL_2017}%
  \BibitemOpen
  \bibfield  {author} {\bibinfo {author} {\bibfnamefont {S.-H.}\ \bibnamefont
  {Baek}}, \bibinfo {author} {\bibfnamefont {S.-H.}\ \bibnamefont {Do}},
  \bibinfo {author} {\bibfnamefont {K.-Y.}\ \bibnamefont {Choi}}, \bibinfo
  {author} {\bibfnamefont {Y.~S.}\ \bibnamefont {Kwon}}, \bibinfo {author}
  {\bibfnamefont {A.~U.~B.}\ \bibnamefont {Wolter}}, \bibinfo {author}
  {\bibfnamefont {S.}~\bibnamefont {Nishimoto}}, \bibinfo {author}
  {\bibfnamefont {J.}~\bibnamefont {van~den Brink}},\ and\ \bibinfo {author}
  {\bibfnamefont {B.}~\bibnamefont {B\"uchner}},\ }\bibfield  {title} {\bibinfo
  {title} {{Evidence for a Field-Induced Quantum Spin Liquid in
  $\ensuremath{\alpha}$-${\mathrm{RuCl}}_{3}$}},\ }\href
  {https://doi.org/10.1103/PhysRevLett.119.037201} {\bibfield  {journal}
  {\bibinfo  {journal} {Phys. Rev. Lett.}\ }\textbf {\bibinfo {volume} {119}},\
  \bibinfo {pages} {037201} (\bibinfo {year} {2017})}\BibitemShut {NoStop}%
\bibitem [{\citenamefont {Banerjee}\ \emph {et~al.}(2018)\citenamefont
  {Banerjee}, \citenamefont {Lampen-Kelley}, \citenamefont {Knolle},
  \citenamefont {Balz}, \citenamefont {Aczel}, \citenamefont {Winn},
  \citenamefont {Liu}, \citenamefont {Pajerowski}, \citenamefont {Yan},
  \citenamefont {Bridges}, \citenamefont {Savici}, \citenamefont {Chakoumakos},
  \citenamefont {Lumsden}, \citenamefont {Tennant}, \citenamefont {Moessner},
  \citenamefont {Mandrus},\ and\ \citenamefont
  {Nagler}}]{banerjee_excitations_field-induced_QSL_2018}%
  \BibitemOpen
  \bibfield  {author} {\bibinfo {author} {\bibfnamefont {A.}~\bibnamefont
  {Banerjee}}, \bibinfo {author} {\bibfnamefont {P.}~\bibnamefont
  {Lampen-Kelley}}, \bibinfo {author} {\bibfnamefont {J.}~\bibnamefont
  {Knolle}}, \bibinfo {author} {\bibfnamefont {C.}~\bibnamefont {Balz}},
  \bibinfo {author} {\bibfnamefont {A.~A.}\ \bibnamefont {Aczel}}, \bibinfo
  {author} {\bibfnamefont {B.}~\bibnamefont {Winn}}, \bibinfo {author}
  {\bibfnamefont {Y.}~\bibnamefont {Liu}}, \bibinfo {author} {\bibfnamefont
  {D.}~\bibnamefont {Pajerowski}}, \bibinfo {author} {\bibfnamefont
  {J.}~\bibnamefont {Yan}}, \bibinfo {author} {\bibfnamefont {C.~A.}\
  \bibnamefont {Bridges}}, \bibinfo {author} {\bibfnamefont {A.~T.}\
  \bibnamefont {Savici}}, \bibinfo {author} {\bibfnamefont {B.~C.}\
  \bibnamefont {Chakoumakos}}, \bibinfo {author} {\bibfnamefont {M.~D.}\
  \bibnamefont {Lumsden}}, \bibinfo {author} {\bibfnamefont {D.~A.}\
  \bibnamefont {Tennant}}, \bibinfo {author} {\bibfnamefont {R.}~\bibnamefont
  {Moessner}}, \bibinfo {author} {\bibfnamefont {D.~G.}\ \bibnamefont
  {Mandrus}},\ and\ \bibinfo {author} {\bibfnamefont {S.~E.}\ \bibnamefont
  {Nagler}},\ }\bibfield  {title} {\bibinfo {title} {{Excitations in the
  field-induced quantum spin liquid state of
  $\ensuremath{\alpha}$-${\mathrm{RuCl}}_{3}$}},\ }\href
  {https://doi.org/10.1038/s41535-018-0079-2} {\bibfield  {journal} {\bibinfo
  {journal} {npj Quantum Materials}\ }\textbf {\bibinfo {volume} {3}},\
  \bibinfo {pages} {8} (\bibinfo {year} {2018})}\BibitemShut {NoStop}%
\bibitem [{\citenamefont {Kasahara}\ \emph {et~al.}(2018)\citenamefont
  {Kasahara}, \citenamefont {Ohnishi}, \citenamefont {Mizukami}, \citenamefont
  {Tanaka}, \citenamefont {Ma}, \citenamefont {Sugii}, \citenamefont {Kurita},
  \citenamefont {Tanaka}, \citenamefont {Nasu}, \citenamefont {Motome},
  \citenamefont {Shibauchi},\ and\ \citenamefont
  {Matsuda}}]{kasahara_majorana_2018}%
  \BibitemOpen
  \bibfield  {author} {\bibinfo {author} {\bibfnamefont {Y.}~\bibnamefont
  {Kasahara}}, \bibinfo {author} {\bibfnamefont {T.}~\bibnamefont {Ohnishi}},
  \bibinfo {author} {\bibfnamefont {Y.}~\bibnamefont {Mizukami}}, \bibinfo
  {author} {\bibfnamefont {O.}~\bibnamefont {Tanaka}}, \bibinfo {author}
  {\bibfnamefont {S.}~\bibnamefont {Ma}}, \bibinfo {author} {\bibfnamefont
  {K.}~\bibnamefont {Sugii}}, \bibinfo {author} {\bibfnamefont
  {N.}~\bibnamefont {Kurita}}, \bibinfo {author} {\bibfnamefont
  {H.}~\bibnamefont {Tanaka}}, \bibinfo {author} {\bibfnamefont
  {J.}~\bibnamefont {Nasu}}, \bibinfo {author} {\bibfnamefont {Y.}~\bibnamefont
  {Motome}}, \bibinfo {author} {\bibfnamefont {T.}~\bibnamefont {Shibauchi}},\
  and\ \bibinfo {author} {\bibfnamefont {Y.}~\bibnamefont {Matsuda}},\
  }\bibfield  {title} {\bibinfo {title} {{Majorana Quantization and
  Half-Integer Thermal Quantum {{Hall}} Effect in a {{Kitaev}} Spin Liquid}},\
  }\href {https://doi.org/10.1038/s41586-018-0274-0} {\bibfield  {journal}
  {\bibinfo  {journal} {Nature}\ }\textbf {\bibinfo {volume} {559}},\ \bibinfo
  {pages} {227} (\bibinfo {year} {2018})}\BibitemShut {NoStop}%
\bibitem [{\citenamefont {Yamashita}\ \emph {et~al.}(2020)\citenamefont
  {Yamashita}, \citenamefont {Gouchi}, \citenamefont {Uwatoko}, \citenamefont
  {Kurita},\ and\ \citenamefont {Tanaka}}]{yamashita_sample_dependence_2020}%
  \BibitemOpen
  \bibfield  {author} {\bibinfo {author} {\bibfnamefont {M.}~\bibnamefont
  {Yamashita}}, \bibinfo {author} {\bibfnamefont {J.}~\bibnamefont {Gouchi}},
  \bibinfo {author} {\bibfnamefont {Y.}~\bibnamefont {Uwatoko}}, \bibinfo
  {author} {\bibfnamefont {N.}~\bibnamefont {Kurita}},\ and\ \bibinfo {author}
  {\bibfnamefont {H.}~\bibnamefont {Tanaka}},\ }\bibfield  {title} {\bibinfo
  {title} {{Sample dependence of half-integer quantized thermal Hall effect in
  the Kitaev spin-liquid candidate
  $\ensuremath{\alpha}\text{\ensuremath{-}}{\mathrm{RuCl}}_{3}$}},\ }\href
  {https://doi.org/10.1103/PhysRevB.102.220404} {\bibfield  {journal} {\bibinfo
   {journal} {Phys. Rev. B}\ }\textbf {\bibinfo {volume} {102}},\ \bibinfo
  {pages} {220404} (\bibinfo {year} {2020})}\BibitemShut {NoStop}%
\bibitem [{\citenamefont {Yokoi}\ \emph {et~al.}(2021)\citenamefont {Yokoi},
  \citenamefont {Ma}, \citenamefont {Kasahara}, \citenamefont {Kasahara},
  \citenamefont {Shibauchi}, \citenamefont {Kurita}, \citenamefont {Tanaka},
  \citenamefont {Nasu}, \citenamefont {Motome}, \citenamefont {Hickey},
  \citenamefont {Trebst},\ and\ \citenamefont
  {Matsuda}}]{yokoi_half-integer_2021}%
  \BibitemOpen
  \bibfield  {author} {\bibinfo {author} {\bibfnamefont {T.}~\bibnamefont
  {Yokoi}}, \bibinfo {author} {\bibfnamefont {S.}~\bibnamefont {Ma}}, \bibinfo
  {author} {\bibfnamefont {Y.}~\bibnamefont {Kasahara}}, \bibinfo {author}
  {\bibfnamefont {S.}~\bibnamefont {Kasahara}}, \bibinfo {author}
  {\bibfnamefont {T.}~\bibnamefont {Shibauchi}}, \bibinfo {author}
  {\bibfnamefont {N.}~\bibnamefont {Kurita}}, \bibinfo {author} {\bibfnamefont
  {H.}~\bibnamefont {Tanaka}}, \bibinfo {author} {\bibfnamefont
  {J.}~\bibnamefont {Nasu}}, \bibinfo {author} {\bibfnamefont {Y.}~\bibnamefont
  {Motome}}, \bibinfo {author} {\bibfnamefont {C.}~\bibnamefont {Hickey}},
  \bibinfo {author} {\bibfnamefont {S.}~\bibnamefont {Trebst}},\ and\ \bibinfo
  {author} {\bibfnamefont {Y.}~\bibnamefont {Matsuda}},\ }\bibfield  {title}
  {\bibinfo {title} {{Half-integer quantized anomalous thermal Hall effect in
  the Kitaev material candidate
  $\ensuremath{\alpha}\text{\ensuremath{-}}{\mathrm{RuCl}}_{3}$}},\ }\href
  {https://doi.org/10.1126/science.aay5551} {\bibfield  {journal} {\bibinfo
  {journal} {Science}\ }\textbf {\bibinfo {volume} {373}},\ \bibinfo {pages}
  {568} (\bibinfo {year} {2021})}\BibitemShut {NoStop}%
\bibitem [{\citenamefont {Bruin}\ \emph {et~al.}(2022)\citenamefont {Bruin},
  \citenamefont {Claus}, \citenamefont {Matsumoto}, \citenamefont {Kurita},
  \citenamefont {Tanaka},\ and\ \citenamefont
  {Takagi}}]{bruin_robustness_2022}%
  \BibitemOpen
  \bibfield  {author} {\bibinfo {author} {\bibfnamefont {J.~A.~N.}\
  \bibnamefont {Bruin}}, \bibinfo {author} {\bibfnamefont {R.~R.}\ \bibnamefont
  {Claus}}, \bibinfo {author} {\bibfnamefont {Y.}~\bibnamefont {Matsumoto}},
  \bibinfo {author} {\bibfnamefont {N.}~\bibnamefont {Kurita}}, \bibinfo
  {author} {\bibfnamefont {H.}~\bibnamefont {Tanaka}},\ and\ \bibinfo {author}
  {\bibfnamefont {H.}~\bibnamefont {Takagi}},\ }\bibfield  {title} {\bibinfo
  {title} {{Robustness of the thermal Hall effect close to half-quantization in
  $\ensuremath{\alpha}\text{\ensuremath{-}}{\mathrm{RuCl}}_{3}$}},\ }\href
  {https://doi.org/10.1038/s41567-021-01501-y} {\bibfield  {journal} {\bibinfo
  {journal} {Nature Physics}\ }\textbf {\bibinfo {volume} {18}},\ \bibinfo
  {pages} {401} (\bibinfo {year} {2022})}\BibitemShut {NoStop}%
\bibitem [{\citenamefont {Ran}\ \emph {et~al.}(2017)\citenamefont {Ran},
  \citenamefont {Wang}, \citenamefont {Wang}, \citenamefont {Dong},
  \citenamefont {Ren}, \citenamefont {Bao}, \citenamefont {Li}, \citenamefont
  {Ma}, \citenamefont {Gan}, \citenamefont {Zhang}, \citenamefont {Park},
  \citenamefont {Deng}, \citenamefont {Danilkin}, \citenamefont {Yu},
  \citenamefont {Li},\ and\ \citenamefont {Wen}}]{Kejing2017_PRL}%
  \BibitemOpen
  \bibfield  {author} {\bibinfo {author} {\bibfnamefont {K.}~\bibnamefont
  {Ran}}, \bibinfo {author} {\bibfnamefont {J.}~\bibnamefont {Wang}}, \bibinfo
  {author} {\bibfnamefont {W.}~\bibnamefont {Wang}}, \bibinfo {author}
  {\bibfnamefont {Z.-Y.}\ \bibnamefont {Dong}}, \bibinfo {author}
  {\bibfnamefont {X.}~\bibnamefont {Ren}}, \bibinfo {author} {\bibfnamefont
  {S.}~\bibnamefont {Bao}}, \bibinfo {author} {\bibfnamefont {S.}~\bibnamefont
  {Li}}, \bibinfo {author} {\bibfnamefont {Z.}~\bibnamefont {Ma}}, \bibinfo
  {author} {\bibfnamefont {Y.}~\bibnamefont {Gan}}, \bibinfo {author}
  {\bibfnamefont {Y.}~\bibnamefont {Zhang}}, \bibinfo {author} {\bibfnamefont
  {J.~T.}\ \bibnamefont {Park}}, \bibinfo {author} {\bibfnamefont
  {G.}~\bibnamefont {Deng}}, \bibinfo {author} {\bibfnamefont {S.}~\bibnamefont
  {Danilkin}}, \bibinfo {author} {\bibfnamefont {S.-L.}\ \bibnamefont {Yu}},
  \bibinfo {author} {\bibfnamefont {J.-X.}\ \bibnamefont {Li}},\ and\ \bibinfo
  {author} {\bibfnamefont {J.}~\bibnamefont {Wen}},\ }\bibfield  {title}
  {\bibinfo {title} {{Spin-Wave Excitations Evidencing the Kitaev Interaction
  in Single Crystalline
  $\ensuremath{\alpha}\text{\ensuremath{-}}{\mathrm{RuCl}}_{3}$}},\ }\href
  {https://doi.org/10.1103/PhysRevLett.118.107203} {\bibfield  {journal}
  {\bibinfo  {journal} {Phys. Rev. Lett.}\ }\textbf {\bibinfo {volume} {118}},\
  \bibinfo {pages} {107203} (\bibinfo {year} {2017})}\BibitemShut {NoStop}%
\bibitem [{\citenamefont {Janssen}\ \emph {et~al.}(2017)\citenamefont
  {Janssen}, \citenamefont {Andrade},\ and\ \citenamefont
  {Vojta}}]{Janssen2017_PRB}%
  \BibitemOpen
  \bibfield  {author} {\bibinfo {author} {\bibfnamefont {L.}~\bibnamefont
  {Janssen}}, \bibinfo {author} {\bibfnamefont {E.~C.}\ \bibnamefont
  {Andrade}},\ and\ \bibinfo {author} {\bibfnamefont {M.}~\bibnamefont
  {Vojta}},\ }\bibfield  {title} {\bibinfo {title} {{Magnetization processes of
  zigzag states on the honeycomb lattice: Identifying spin models for
  $\ensuremath{\alpha}\text{\ensuremath{-}}{\mathrm{RuCl}}_{3}$ and
  ${\mathrm{Na}}_{2}{\mathrm{IrO}}_{3}$}},\ }\href
  {https://doi.org/10.1103/PhysRevB.96.064430} {\bibfield  {journal} {\bibinfo
  {journal} {Phys. Rev. B}\ }\textbf {\bibinfo {volume} {96}},\ \bibinfo
  {pages} {064430} (\bibinfo {year} {2017})}\BibitemShut {NoStop}%
\bibitem [{\citenamefont {Wang}\ \emph {et~al.}(2017)\citenamefont {Wang},
  \citenamefont {Dong}, \citenamefont {Yu},\ and\ \citenamefont
  {Li}}]{wang_PRB_2017}%
  \BibitemOpen
  \bibfield  {author} {\bibinfo {author} {\bibfnamefont {W.}~\bibnamefont
  {Wang}}, \bibinfo {author} {\bibfnamefont {Z.-Y.}\ \bibnamefont {Dong}},
  \bibinfo {author} {\bibfnamefont {S.-L.}\ \bibnamefont {Yu}},\ and\ \bibinfo
  {author} {\bibfnamefont {J.-X.}\ \bibnamefont {Li}},\ }\bibfield  {title}
  {\bibinfo {title} {{Theoretical investigation of magnetic dynamics in
  $\ensuremath{\alpha}\ensuremath{-}{\mathrm{RuCl}}_{3}$}},\ }\href
  {https://doi.org/10.1103/PhysRevB.96.115103} {\bibfield  {journal} {\bibinfo
  {journal} {Phys. Rev. B}\ }\textbf {\bibinfo {volume} {96}},\ \bibinfo
  {pages} {115103} (\bibinfo {year} {2017})}\BibitemShut {NoStop}%
\bibitem [{\citenamefont {Winter}\ \emph
  {et~al.}(2017{\natexlab{b}})\citenamefont {Winter}, \citenamefont {Riedl},
  \citenamefont {Maksimov}, \citenamefont {Chernyshev}, \citenamefont
  {Honecker},\ and\ \citenamefont {Valenti}}]{Winter2017_NatMat}%
  \BibitemOpen
  \bibfield  {author} {\bibinfo {author} {\bibfnamefont {S.~M.}\ \bibnamefont
  {Winter}}, \bibinfo {author} {\bibfnamefont {K.}~\bibnamefont {Riedl}},
  \bibinfo {author} {\bibfnamefont {P.~A.}\ \bibnamefont {Maksimov}}, \bibinfo
  {author} {\bibfnamefont {A.~L.}\ \bibnamefont {Chernyshev}}, \bibinfo
  {author} {\bibfnamefont {A.}~\bibnamefont {Honecker}},\ and\ \bibinfo
  {author} {\bibfnamefont {R.}~\bibnamefont {Valenti}},\ }\bibfield  {title}
  {\bibinfo {title} {{Breakdown of magnons in a strongly spin-orbital coupled
  magnet}},\ }\href {https://www.nature.com/articles/s41467-017-01177-0}
  {\bibfield  {journal} {\bibinfo  {journal} {Nature Communications}\ }\textbf
  {\bibinfo {volume} {8}},\ \bibinfo {pages} {1152} (\bibinfo {year}
  {2017}{\natexlab{b}})}\BibitemShut {NoStop}%
\bibitem [{\citenamefont {Wang}\ \emph {et~al.}(2019)\citenamefont {Wang},
  \citenamefont {Normand},\ and\ \citenamefont {Liu}}]{wang_JKG_VMC_2019}%
  \BibitemOpen
  \bibfield  {author} {\bibinfo {author} {\bibfnamefont {J.}~\bibnamefont
  {Wang}}, \bibinfo {author} {\bibfnamefont {B.}~\bibnamefont {Normand}},\ and\
  \bibinfo {author} {\bibfnamefont {Z.-X.}\ \bibnamefont {Liu}},\ }\bibfield
  {title} {\bibinfo {title} {{One Proximate Kitaev Spin Liquid in the
  $K\text{\ensuremath{-}}J\text{\ensuremath{-}}\mathrm{\ensuremath{\Gamma}}$
  Model on the Honeycomb Lattice}},\ }\href
  {https://doi.org/10.1103/PhysRevLett.123.197201} {\bibfield  {journal}
  {\bibinfo  {journal} {Phys. Rev. Lett.}\ }\textbf {\bibinfo {volume} {123}},\
  \bibinfo {pages} {197201} (\bibinfo {year} {2019})}\BibitemShut {NoStop}%
\bibitem [{\citenamefont {Lee}\ \emph {et~al.}(2019)\citenamefont {Lee},
  \citenamefont {Kaneko}, \citenamefont {Chern}, \citenamefont {Okubo},
  \citenamefont {Yamaji}, \citenamefont {Kawashima},\ and\ \citenamefont
  {Kim}}]{lee_magneticfield_2019}%
  \BibitemOpen
  \bibfield  {author} {\bibinfo {author} {\bibfnamefont {H.-Y.}\ \bibnamefont
  {Lee}}, \bibinfo {author} {\bibfnamefont {R.}~\bibnamefont {Kaneko}},
  \bibinfo {author} {\bibfnamefont {L.~E.}\ \bibnamefont {Chern}}, \bibinfo
  {author} {\bibfnamefont {T.}~\bibnamefont {Okubo}}, \bibinfo {author}
  {\bibfnamefont {Y.}~\bibnamefont {Yamaji}}, \bibinfo {author} {\bibfnamefont
  {N.}~\bibnamefont {Kawashima}},\ and\ \bibinfo {author} {\bibfnamefont
  {Y.~B.}\ \bibnamefont {Kim}},\ }\bibfield  {title} {\bibinfo {title}
  {{Magnetic-Field Induced Quantum Phases in Tensor Network Study of Kitaev
  Magnets}},\ }\href@noop {} {\bibfield  {journal} {\bibinfo  {journal}
  {arXiv:1908.07671 [cond-mat]}\ } (\bibinfo {year} {2019})}\BibitemShut
  {NoStop}%
\bibitem [{\citenamefont {Buessen}\ and\ \citenamefont
  {Kim}(2021)}]{buessen_KGamma_FRG_2021}%
  \BibitemOpen
  \bibfield  {author} {\bibinfo {author} {\bibfnamefont {F.~L.}\ \bibnamefont
  {Buessen}}\ and\ \bibinfo {author} {\bibfnamefont {Y.~B.}\ \bibnamefont
  {Kim}},\ }\bibfield  {title} {\bibinfo {title} {{Functional renormalization
  group study of the Kitaev-$\mathrm{\ensuremath{\Gamma}}$ model on the
  honeycomb lattice and emergent incommensurate magnetic correlations}},\
  }\href {https://doi.org/10.1103/PhysRevB.103.184407} {\bibfield  {journal}
  {\bibinfo  {journal} {Phys. Rev. B}\ }\textbf {\bibinfo {volume} {103}},\
  \bibinfo {pages} {184407} (\bibinfo {year} {2021})}\BibitemShut {NoStop}%
\bibitem [{\citenamefont {Gohlke}\ \emph {et~al.}(2018)\citenamefont {Gohlke},
  \citenamefont {Wachtel}, \citenamefont {Yamaji}, \citenamefont {Pollmann},\
  and\ \citenamefont {Kim}}]{GohlkePRB2018}%
  \BibitemOpen
  \bibfield  {author} {\bibinfo {author} {\bibfnamefont {M.}~\bibnamefont
  {Gohlke}}, \bibinfo {author} {\bibfnamefont {G.}~\bibnamefont {Wachtel}},
  \bibinfo {author} {\bibfnamefont {Y.}~\bibnamefont {Yamaji}}, \bibinfo
  {author} {\bibfnamefont {F.}~\bibnamefont {Pollmann}},\ and\ \bibinfo
  {author} {\bibfnamefont {Y.~B.}\ \bibnamefont {Kim}},\ }\bibfield  {title}
  {\bibinfo {title} {{Quantum spin liquid signatures in Kitaev-like frustrated
  magnets}},\ }\href {https://doi.org/10.1103/PhysRevB.97.075126} {\bibfield
  {journal} {\bibinfo  {journal} {Phys. Rev. B}\ }\textbf {\bibinfo {volume}
  {97}},\ \bibinfo {pages} {075126} (\bibinfo {year} {2018})}\BibitemShut
  {NoStop}%
\bibitem [{\citenamefont {Gohlke}\ \emph {et~al.}(2020)\citenamefont {Gohlke},
  \citenamefont {Chern}, \citenamefont {Kee},\ and\ \citenamefont
  {Kim}}]{gohlke_lattice-nematic_2020}%
  \BibitemOpen
  \bibfield  {author} {\bibinfo {author} {\bibfnamefont {M.}~\bibnamefont
  {Gohlke}}, \bibinfo {author} {\bibfnamefont {L.~E.}\ \bibnamefont {Chern}},
  \bibinfo {author} {\bibfnamefont {H.-Y.}\ \bibnamefont {Kee}},\ and\ \bibinfo
  {author} {\bibfnamefont {Y.~B.}\ \bibnamefont {Kim}},\ }\bibfield  {title}
  {\bibinfo {title} {{Emergence of nematic paramagnet via quantum
  order-by-disorder and pseudo-Goldstone modes in Kitaev magnets}},\ }\href
  {https://doi.org/10.1103/PhysRevResearch.2.043023} {\bibfield  {journal}
  {\bibinfo  {journal} {Phys. Rev. Research}\ }\textbf {\bibinfo {volume}
  {2}},\ \bibinfo {pages} {043023} (\bibinfo {year} {2020})}\BibitemShut
  {NoStop}%
\bibitem [{\citenamefont {Johnson}\ \emph {et~al.}(2015)\citenamefont
  {Johnson}, \citenamefont {Williams}, \citenamefont {Haghighirad},
  \citenamefont {Singleton}, \citenamefont {Zapf}, \citenamefont {Manuel},
  \citenamefont {Mazin}, \citenamefont {Li}, \citenamefont {Jeschke},
  \citenamefont {Valent\'{\i}},\ and\ \citenamefont
  {Coldea}}]{johnson_monoclinic_2015}%
  \BibitemOpen
  \bibfield  {author} {\bibinfo {author} {\bibfnamefont {R.~D.}\ \bibnamefont
  {Johnson}}, \bibinfo {author} {\bibfnamefont {S.~C.}\ \bibnamefont
  {Williams}}, \bibinfo {author} {\bibfnamefont {A.~A.}\ \bibnamefont
  {Haghighirad}}, \bibinfo {author} {\bibfnamefont {J.}~\bibnamefont
  {Singleton}}, \bibinfo {author} {\bibfnamefont {V.}~\bibnamefont {Zapf}},
  \bibinfo {author} {\bibfnamefont {P.}~\bibnamefont {Manuel}}, \bibinfo
  {author} {\bibfnamefont {I.~I.}\ \bibnamefont {Mazin}}, \bibinfo {author}
  {\bibfnamefont {Y.}~\bibnamefont {Li}}, \bibinfo {author} {\bibfnamefont
  {H.~O.}\ \bibnamefont {Jeschke}}, \bibinfo {author} {\bibfnamefont
  {R.}~\bibnamefont {Valent\'{\i}}},\ and\ \bibinfo {author} {\bibfnamefont
  {R.}~\bibnamefont {Coldea}},\ }\bibfield  {title} {\bibinfo {title}
  {{Monoclinic crystal structure of
  $\ensuremath{\alpha}\ensuremath{-}{\mathrm{RuCl}}_{3}$ and the zigzag
  antiferromagnetic ground state}},\ }\href
  {https://doi.org/10.1103/PhysRevB.92.235119} {\bibfield  {journal} {\bibinfo
  {journal} {Phys. Rev. B}\ }\textbf {\bibinfo {volume} {92}},\ \bibinfo
  {pages} {235119} (\bibinfo {year} {2015})}\BibitemShut {NoStop}%
\bibitem [{\citenamefont {Cao}\ \emph {et~al.}(2016)\citenamefont {Cao},
  \citenamefont {Banerjee}, \citenamefont {Yan}, \citenamefont {Bridges},
  \citenamefont {Lumsden}, \citenamefont {Mandrus}, \citenamefont {Tennant},
  \citenamefont {Chakoumakos},\ and\ \citenamefont
  {Nagler}}]{cao_low-T_structure_2016}%
  \BibitemOpen
  \bibfield  {author} {\bibinfo {author} {\bibfnamefont {H.~B.}\ \bibnamefont
  {Cao}}, \bibinfo {author} {\bibfnamefont {A.}~\bibnamefont {Banerjee}},
  \bibinfo {author} {\bibfnamefont {J.-Q.}\ \bibnamefont {Yan}}, \bibinfo
  {author} {\bibfnamefont {C.~A.}\ \bibnamefont {Bridges}}, \bibinfo {author}
  {\bibfnamefont {M.~D.}\ \bibnamefont {Lumsden}}, \bibinfo {author}
  {\bibfnamefont {D.~G.}\ \bibnamefont {Mandrus}}, \bibinfo {author}
  {\bibfnamefont {D.~A.}\ \bibnamefont {Tennant}}, \bibinfo {author}
  {\bibfnamefont {B.~C.}\ \bibnamefont {Chakoumakos}},\ and\ \bibinfo {author}
  {\bibfnamefont {S.~E.}\ \bibnamefont {Nagler}},\ }\bibfield  {title}
  {\bibinfo {title} {{Low-temperature crystal and magnetic structure of
  $\ensuremath{\alpha}\ensuremath{-}{\mathrm{RuCl}}_{3}$}},\ }\href
  {https://doi.org/10.1103/PhysRevB.93.134423} {\bibfield  {journal} {\bibinfo
  {journal} {Phys. Rev. B}\ }\textbf {\bibinfo {volume} {93}},\ \bibinfo
  {pages} {134423} (\bibinfo {year} {2016})}\BibitemShut {NoStop}%
\bibitem [{\citenamefont {Janssen}\ \emph {et~al.}(2020)\citenamefont
  {Janssen}, \citenamefont {Koch},\ and\ \citenamefont
  {Vojta}}]{janssen_magnon_dispersion_2020}%
  \BibitemOpen
  \bibfield  {author} {\bibinfo {author} {\bibfnamefont {L.}~\bibnamefont
  {Janssen}}, \bibinfo {author} {\bibfnamefont {S.}~\bibnamefont {Koch}},\ and\
  \bibinfo {author} {\bibfnamefont {M.}~\bibnamefont {Vojta}},\ }\bibfield
  {title} {\bibinfo {title} {{Magnon dispersion and dynamic spin response in
  three-dimensional spin models for
  $\ensuremath{\alpha}\text{\ensuremath{-}}{\mathrm{RuCl}}_{3}$}},\ }\href
  {https://doi.org/10.1103/PhysRevB.101.174444} {\bibfield  {journal} {\bibinfo
   {journal} {Phys. Rev. B}\ }\textbf {\bibinfo {volume} {101}},\ \bibinfo
  {pages} {174444} (\bibinfo {year} {2020})}\BibitemShut {NoStop}%
\bibitem [{\citenamefont {Kocsis}\ \emph {et~al.}(2022)\citenamefont {Kocsis},
  \citenamefont {Kaib}, \citenamefont {Riedl}, \citenamefont {Gass},
  \citenamefont {Lampen-Kelley}, \citenamefont {Mandrus}, \citenamefont
  {Nagler}, \citenamefont {P\'erez}, \citenamefont {Nielsch}, \citenamefont
  {B\"uchner}, \citenamefont {Wolter},\ and\ \citenamefont
  {Valent\'{\i}}}]{vilmos_magnetoelastic_coupling_2022}%
  \BibitemOpen
  \bibfield  {author} {\bibinfo {author} {\bibfnamefont {V.}~\bibnamefont
  {Kocsis}}, \bibinfo {author} {\bibfnamefont {D.~A.~S.}\ \bibnamefont {Kaib}},
  \bibinfo {author} {\bibfnamefont {K.}~\bibnamefont {Riedl}}, \bibinfo
  {author} {\bibfnamefont {S.}~\bibnamefont {Gass}}, \bibinfo {author}
  {\bibfnamefont {P.}~\bibnamefont {Lampen-Kelley}}, \bibinfo {author}
  {\bibfnamefont {D.~G.}\ \bibnamefont {Mandrus}}, \bibinfo {author}
  {\bibfnamefont {S.~E.}\ \bibnamefont {Nagler}}, \bibinfo {author}
  {\bibfnamefont {N.}~\bibnamefont {P\'erez}}, \bibinfo {author} {\bibfnamefont
  {K.}~\bibnamefont {Nielsch}}, \bibinfo {author} {\bibfnamefont
  {B.}~\bibnamefont {B\"uchner}}, \bibinfo {author} {\bibfnamefont {A.~U.~B.}\
  \bibnamefont {Wolter}},\ and\ \bibinfo {author} {\bibfnamefont
  {R.}~\bibnamefont {Valent\'{\i}}},\ }\bibfield  {title} {\bibinfo {title}
  {{Magnetoelastic coupling anisotropy in the Kitaev material
  $\ensuremath{\alpha}\text{\ensuremath{-}}\mathrm{Ru}{\mathrm{Cl}}_{3}$}},\
  }\href {https://doi.org/10.1103/PhysRevB.105.094410} {\bibfield  {journal}
  {\bibinfo  {journal} {Phys. Rev. B}\ }\textbf {\bibinfo {volume} {105}},\
  \bibinfo {pages} {094410} (\bibinfo {year} {2022})}\BibitemShut {NoStop}%
\bibitem [{\citenamefont {Kaib}\ \emph {et~al.}(2021)\citenamefont {Kaib},
  \citenamefont {Biswas}, \citenamefont {Riedl}, \citenamefont {Winter},\ and\
  \citenamefont {Valent\'{\i}}}]{kaib_magnetoelastic_coupling_2021}%
  \BibitemOpen
  \bibfield  {author} {\bibinfo {author} {\bibfnamefont {D.~A.~S.}\
  \bibnamefont {Kaib}}, \bibinfo {author} {\bibfnamefont {S.}~\bibnamefont
  {Biswas}}, \bibinfo {author} {\bibfnamefont {K.}~\bibnamefont {Riedl}},
  \bibinfo {author} {\bibfnamefont {S.~M.}\ \bibnamefont {Winter}},\ and\
  \bibinfo {author} {\bibfnamefont {R.}~\bibnamefont {Valent\'{\i}}},\
  }\bibfield  {title} {\bibinfo {title} {{Magnetoelastic coupling and effects
  of uniaxial strain in $\ensuremath{\alpha}\ensuremath{-}{\mathrm{RuCl}}_{3}$
  from first principles}},\ }\href
  {https://doi.org/10.1103/PhysRevB.103.L140402} {\bibfield  {journal}
  {\bibinfo  {journal} {Phys. Rev. B}\ }\textbf {\bibinfo {volume} {103}},\
  \bibinfo {pages} {L140402} (\bibinfo {year} {2021})}\BibitemShut {NoStop}%
\bibitem [{\citenamefont {Wang}\ and\ \citenamefont
  {Liu}(2020)}]{wang_anisoKG_VMC_2020}%
  \BibitemOpen
  \bibfield  {author} {\bibinfo {author} {\bibfnamefont {J.}~\bibnamefont
  {Wang}}\ and\ \bibinfo {author} {\bibfnamefont {Z.-X.}\ \bibnamefont {Liu}},\
  }\bibfield  {title} {\bibinfo {title} {{Symmetry-protected gapless spin
  liquids on the strained honeycomb lattice}},\ }\href
  {https://doi.org/10.1103/PhysRevB.102.094416} {\bibfield  {journal} {\bibinfo
   {journal} {Phys. Rev. B}\ }\textbf {\bibinfo {volume} {102}},\ \bibinfo
  {pages} {094416} (\bibinfo {year} {2020})}\BibitemShut {NoStop}%
\bibitem [{\citenamefont {Yang}\ \emph
  {et~al.}(2020{\natexlab{a}})\citenamefont {Yang}, \citenamefont {Nocera},
  \citenamefont {Tummuru}, \citenamefont {Kee},\ and\ \citenamefont
  {Affleck}}]{yang_KGchain_2020}%
  \BibitemOpen
  \bibfield  {author} {\bibinfo {author} {\bibfnamefont {W.}~\bibnamefont
  {Yang}}, \bibinfo {author} {\bibfnamefont {A.}~\bibnamefont {Nocera}},
  \bibinfo {author} {\bibfnamefont {T.}~\bibnamefont {Tummuru}}, \bibinfo
  {author} {\bibfnamefont {H.-Y.}\ \bibnamefont {Kee}},\ and\ \bibinfo {author}
  {\bibfnamefont {I.}~\bibnamefont {Affleck}},\ }\bibfield  {title} {\bibinfo
  {title} {{Phase Diagram of the Spin-$1/2$ Kitaev-Gamma Chain and Emergent
  SU(2) Symmetry}},\ }\href {https://doi.org/10.1103/PhysRevLett.124.147205}
  {\bibfield  {journal} {\bibinfo  {journal} {Phys. Rev. Lett.}\ }\textbf
  {\bibinfo {volume} {124}},\ \bibinfo {pages} {147205} (\bibinfo {year}
  {2020}{\natexlab{a}})}\BibitemShut {NoStop}%
\bibitem [{\citenamefont {Giamarchi}(2003)}]{giamarchi_book}%
  \BibitemOpen
  \bibfield  {author} {\bibinfo {author} {\bibfnamefont {T.}~\bibnamefont
  {Giamarchi}},\ }\href {https://books.google.co.jp/books?id=0CGVxiyUZYYC}
  {\emph {\bibinfo {title} {{Quantum Physics in One Dimension}}}},\
  International Series of Monographs on Physics\ (\bibinfo  {publisher}
  {Clarendon Press},\ \bibinfo {year} {2003})\BibitemShut {NoStop}%
\bibitem [{\citenamefont {Schulz}(1996)}]{schulz_dynamics_of_1996}%
  \BibitemOpen
  \bibfield  {author} {\bibinfo {author} {\bibfnamefont {H.~J.}\ \bibnamefont
  {Schulz}},\ }\bibfield  {title} {\bibinfo {title} {{Dynamics of Coupled
  Quantum Spin Chains}},\ }\href {https://doi.org/10.1103/PhysRevLett.77.2790}
  {\bibfield  {journal} {\bibinfo  {journal} {Phys. Rev. Lett.}\ }\textbf
  {\bibinfo {volume} {77}},\ \bibinfo {pages} {2790} (\bibinfo {year}
  {1996})}\BibitemShut {NoStop}%
\bibitem [{\citenamefont {Emery}\ \emph {et~al.}(2000)\citenamefont {Emery},
  \citenamefont {Fradkin}, \citenamefont {Kivelson},\ and\ \citenamefont
  {Lubensky}}]{emery_smectic_metal_state_2000}%
  \BibitemOpen
  \bibfield  {author} {\bibinfo {author} {\bibfnamefont {V.~J.}\ \bibnamefont
  {Emery}}, \bibinfo {author} {\bibfnamefont {E.}~\bibnamefont {Fradkin}},
  \bibinfo {author} {\bibfnamefont {S.~A.}\ \bibnamefont {Kivelson}},\ and\
  \bibinfo {author} {\bibfnamefont {T.~C.}\ \bibnamefont {Lubensky}},\
  }\bibfield  {title} {\bibinfo {title} {{Quantum Theory of the Smectic Metal
  State in Stripe Phases}},\ }\href
  {https://doi.org/10.1103/PhysRevLett.85.2160} {\bibfield  {journal} {\bibinfo
   {journal} {Phys. Rev. Lett.}\ }\textbf {\bibinfo {volume} {85}},\ \bibinfo
  {pages} {2160} (\bibinfo {year} {2000})}\BibitemShut {NoStop}%
\bibitem [{\citenamefont {Vishwanath}\ and\ \citenamefont
  {Carpentier}(2001)}]{vishwanath_nonfermiliquid_2001}%
  \BibitemOpen
  \bibfield  {author} {\bibinfo {author} {\bibfnamefont {A.}~\bibnamefont
  {Vishwanath}}\ and\ \bibinfo {author} {\bibfnamefont {D.}~\bibnamefont
  {Carpentier}},\ }\bibfield  {title} {\bibinfo {title} {{Two-Dimensional
  Anisotropic Non-Fermi-Liquid Phase of Coupled Luttinger Liquids}},\ }\href
  {https://doi.org/10.1103/PhysRevLett.86.676} {\bibfield  {journal} {\bibinfo
  {journal} {Phys. Rev. Lett.}\ }\textbf {\bibinfo {volume} {86}},\ \bibinfo
  {pages} {676} (\bibinfo {year} {2001})}\BibitemShut {NoStop}%
\bibitem [{\citenamefont {Mukhopadhyay}\ \emph {et~al.}(2001)\citenamefont
  {Mukhopadhyay}, \citenamefont {Kane},\ and\ \citenamefont
  {Lubensky}}]{mukhopadhyay_sliding_luttinger_2001}%
  \BibitemOpen
  \bibfield  {author} {\bibinfo {author} {\bibfnamefont {R.}~\bibnamefont
  {Mukhopadhyay}}, \bibinfo {author} {\bibfnamefont {C.~L.}\ \bibnamefont
  {Kane}},\ and\ \bibinfo {author} {\bibfnamefont {T.~C.}\ \bibnamefont
  {Lubensky}},\ }\bibfield  {title} {\bibinfo {title} {{Sliding Luttinger
  liquid phases}},\ }\href {https://doi.org/10.1103/PhysRevB.64.045120}
  {\bibfield  {journal} {\bibinfo  {journal} {Phys. Rev. B}\ }\textbf {\bibinfo
  {volume} {64}},\ \bibinfo {pages} {045120} (\bibinfo {year}
  {2001})}\BibitemShut {NoStop}%
\bibitem [{\citenamefont {Yang}\ \emph
  {et~al.}(2020{\natexlab{b}})\citenamefont {Yang}, \citenamefont {Nocera},\
  and\ \citenamefont {Affleck}}]{yang_JKGchain_2020}%
  \BibitemOpen
  \bibfield  {author} {\bibinfo {author} {\bibfnamefont {W.}~\bibnamefont
  {Yang}}, \bibinfo {author} {\bibfnamefont {A.}~\bibnamefont {Nocera}},\ and\
  \bibinfo {author} {\bibfnamefont {I.}~\bibnamefont {Affleck}},\ }\bibfield
  {title} {\bibinfo {title} {{Comprehensive study of the phase diagram of the
  spin-$\frac{1}{2}$ Kitaev-Heisenberg-Gamma chain}},\ }\href
  {https://doi.org/10.1103/PhysRevResearch.2.033268} {\bibfield  {journal}
  {\bibinfo  {journal} {Phys. Rev. Research}\ }\textbf {\bibinfo {volume}
  {2}},\ \bibinfo {pages} {033268} (\bibinfo {year}
  {2020}{\natexlab{b}})}\BibitemShut {NoStop}%
\bibitem [{\citenamefont {White}(1992)}]{white_dmrg_1992}%
  \BibitemOpen
  \bibfield  {author} {\bibinfo {author} {\bibfnamefont {S.~R.}\ \bibnamefont
  {White}},\ }\bibfield  {title} {\bibinfo {title} {{Density Matrix Formulation
  for Quantum Renormalization Groups}},\ }\href
  {https://doi.org/10.1103/PhysRevLett.69.2863} {\bibfield  {journal} {\bibinfo
   {journal} {Phys. Rev. Lett.}\ }\textbf {\bibinfo {volume} {69}},\ \bibinfo
  {pages} {2863} (\bibinfo {year} {1992})}\BibitemShut {NoStop}%
\bibitem [{\citenamefont {McCulloch}(2008)}]{mcculloch_infinite_2008}%
  \BibitemOpen
  \bibfield  {author} {\bibinfo {author} {\bibfnamefont {I.~P.}\ \bibnamefont
  {McCulloch}},\ }\bibfield  {title} {\bibinfo {title} {{Infinite Size Density
  Matrix Renormalization Group, Revisited}}} (\bibinfo {year}
  {2008})\BibitemShut {NoStop}%
\bibitem [{\citenamefont {Phien}\ \emph {et~al.}(2012)\citenamefont {Phien},
  \citenamefont {Vidal},\ and\ \citenamefont
  {McCulloch}}]{phien_infinite_2012}%
  \BibitemOpen
  \bibfield  {author} {\bibinfo {author} {\bibfnamefont {H.~N.}\ \bibnamefont
  {Phien}}, \bibinfo {author} {\bibfnamefont {G.}~\bibnamefont {Vidal}},\ and\
  \bibinfo {author} {\bibfnamefont {I.~P.}\ \bibnamefont {McCulloch}},\
  }\bibfield  {title} {\bibinfo {title} {{Infinite Boundary Conditions for
  Matrix Product State Calculations}},\ }\href
  {https://doi.org/10.1103/PhysRevB.86.245107} {\bibfield  {journal} {\bibinfo
  {journal} {Phys. Rev. B}\ }\textbf {\bibinfo {volume} {86}},\ \bibinfo
  {pages} {245107} (\bibinfo {year} {2012})}\BibitemShut {NoStop}%
\bibitem [{\citenamefont {Zaletel}\ \emph {et~al.}(2015)\citenamefont
  {Zaletel}, \citenamefont {Mong}, \citenamefont {Karrasch}, \citenamefont
  {Moore},\ and\ \citenamefont {Pollmann}}]{zaletel_tmpo_2015}%
  \BibitemOpen
  \bibfield  {author} {\bibinfo {author} {\bibfnamefont {M.~P.}\ \bibnamefont
  {Zaletel}}, \bibinfo {author} {\bibfnamefont {R.~S.~K.}\ \bibnamefont
  {Mong}}, \bibinfo {author} {\bibfnamefont {C.}~\bibnamefont {Karrasch}},
  \bibinfo {author} {\bibfnamefont {J.~E.}\ \bibnamefont {Moore}},\ and\
  \bibinfo {author} {\bibfnamefont {F.}~\bibnamefont {Pollmann}},\ }\bibfield
  {title} {\bibinfo {title} {{Time-Evolving a Matrix Product State with
  Long-Ranged Interactions}},\ }\href
  {https://doi.org/10.1103/PhysRevB.91.165112} {\bibfield  {journal} {\bibinfo
  {journal} {Phys. Rev. B}\ }\textbf {\bibinfo {volume} {91}},\ \bibinfo
  {pages} {165112} (\bibinfo {year} {2015})}\BibitemShut {NoStop}%
\bibitem [{\citenamefont {Gohlke}\ \emph {et~al.}(2017)\citenamefont {Gohlke},
  \citenamefont {Verresen}, \citenamefont {Moessner},\ and\ \citenamefont
  {Pollmann}}]{gohlke_dynamics_JK_2017}%
  \BibitemOpen
  \bibfield  {author} {\bibinfo {author} {\bibfnamefont {M.}~\bibnamefont
  {Gohlke}}, \bibinfo {author} {\bibfnamefont {R.}~\bibnamefont {Verresen}},
  \bibinfo {author} {\bibfnamefont {R.}~\bibnamefont {Moessner}},\ and\
  \bibinfo {author} {\bibfnamefont {F.}~\bibnamefont {Pollmann}},\ }\bibfield
  {title} {\bibinfo {title} {{Dynamics of the Kitaev-Heisenberg Model}},\
  }\href {https://doi.org/10.1103/PhysRevLett.119.157203} {\bibfield  {journal}
  {\bibinfo  {journal} {Phys. Rev. Lett.}\ }\textbf {\bibinfo {volume} {119}},\
  \bibinfo {pages} {157203} (\bibinfo {year} {2017})}\BibitemShut {NoStop}%
\bibitem [{\citenamefont {Yamada}\ \emph {et~al.}(2020)\citenamefont {Yamada},
  \citenamefont {Suzuki},\ and\ \citenamefont {Suga}}]{YamadaT2020}%
  \BibitemOpen
  \bibfield  {author} {\bibinfo {author} {\bibfnamefont {T.}~\bibnamefont
  {Yamada}}, \bibinfo {author} {\bibfnamefont {T.}~\bibnamefont {Suzuki}},\
  and\ \bibinfo {author} {\bibfnamefont {S.-i.}\ \bibnamefont {Suga}},\
  }\bibfield  {title} {\bibinfo {title} {{Ground-state properties of the
  $K\ensuremath{-}\mathrm{\ensuremath{\Gamma}}$ model on a honeycomb
  lattice}},\ }\href {https://doi.org/10.1103/PhysRevB.102.024415} {\bibfield
  {journal} {\bibinfo  {journal} {Phys. Rev. B}\ }\textbf {\bibinfo {volume}
  {102}},\ \bibinfo {pages} {024415} (\bibinfo {year} {2020})}\BibitemShut
  {NoStop}%
\bibitem [{\citenamefont {Andreev}\ and\ \citenamefont
  {Grishchuk}(1984)}]{andreev_spin_1984}%
  \BibitemOpen
  \bibfield  {author} {\bibinfo {author} {\bibfnamefont {A.}~\bibnamefont
  {Andreev}}\ and\ \bibinfo {author} {\bibfnamefont {I.}~\bibnamefont
  {Grishchuk}},\ }\bibfield  {title} {\bibinfo {title} {Spin nematics},\
  }\href@noop {} {\bibfield  {journal} {\bibinfo  {journal} {Sov. Phys. JETP}\
  }\textbf {\bibinfo {volume} {60}},\ \bibinfo {pages} {267} (\bibinfo {year}
  {1984})}\BibitemShut {NoStop}%
\bibitem [{\citenamefont {Gohlke}\ \emph {et~al.}()\citenamefont {Gohlke},
  \citenamefont {Pelayo},\ and\ \citenamefont {Suzuki}}]{long_paper}%
  \BibitemOpen
  \bibfield  {author} {\bibinfo {author} {\bibfnamefont {M.}~\bibnamefont
  {Gohlke}}, \bibinfo {author} {\bibfnamefont {J.~C.}\ \bibnamefont {Pelayo}},\
  and\ \bibinfo {author} {\bibfnamefont {T.}~\bibnamefont {Suzuki}},\ }\bibinfo
  {title} {(in preparation)}\BibitemShut {NoStop}%
\bibitem [{\citenamefont {Yang}\ \emph {et~al.}(2022)\citenamefont {Yang},
  \citenamefont {Nocera}, \citenamefont {Xu}, \citenamefont {Kee},\ and\
  \citenamefont {Affleck}}]{yang_coupled_JKGchains_2022}%
  \BibitemOpen
\bibfield  {title} {  }\bibfield  {author} {\bibinfo {author} {\bibfnamefont
  {W.}~\bibnamefont {Yang}}, \bibinfo {author} {\bibfnamefont {A.}~\bibnamefont
  {Nocera}}, \bibinfo {author} {\bibfnamefont {C.}~\bibnamefont {Xu}}, \bibinfo
  {author} {\bibfnamefont {H.-Y.}\ \bibnamefont {Kee}},\ and\ \bibinfo {author}
  {\bibfnamefont {I.}~\bibnamefont {Affleck}},\ }\bibfield  {title} {\bibinfo
  {title} {{Counter-rotating spiral, zigzag, and 120$^\circ$ orders from
  coupled-chain analysis of Kitaev-Gamma-Heisenberg model, and relations to
  honeycomb iridates}},\ }\Eprint {https://arxiv.org/abs/2207.02188}
  {arXiv:2207.02188}  (\bibinfo {year} {2022})\BibitemShut {NoStop}%
\bibitem [{\citenamefont {Zhang}\ \emph {et~al.}(2021)\citenamefont {Zhang},
  \citenamefont {Hal\'asz}, \citenamefont {Zhu},\ and\ \citenamefont
  {Batista}}]{ZhangBatistaPRB2021}%
  \BibitemOpen
  \bibfield  {author} {\bibinfo {author} {\bibfnamefont {S.-S.}\ \bibnamefont
  {Zhang}}, \bibinfo {author} {\bibfnamefont {G.~B.}\ \bibnamefont {Hal\'asz}},
  \bibinfo {author} {\bibfnamefont {W.}~\bibnamefont {Zhu}},\ and\ \bibinfo
  {author} {\bibfnamefont {C.~D.}\ \bibnamefont {Batista}},\ }\bibfield
  {title} {\bibinfo {title} {{Variational study of the Kitaev-Heisenberg-Gamma
  model}},\ }\href {https://doi.org/10.1103/PhysRevB.104.014411} {\bibfield
  {journal} {\bibinfo  {journal} {Phys. Rev. B}\ }\textbf {\bibinfo {volume}
  {104}},\ \bibinfo {pages} {014411} (\bibinfo {year} {2021})}\BibitemShut
  {NoStop}%
\bibitem [{\citenamefont {Rams}\ \emph {et~al.}(2018)\citenamefont {Rams},
  \citenamefont {Czarnik},\ and\ \citenamefont
  {Cincio}}]{rams_precise_extrapolation_2018}%
  \BibitemOpen
  \bibfield  {author} {\bibinfo {author} {\bibfnamefont {M.~M.}\ \bibnamefont
  {Rams}}, \bibinfo {author} {\bibfnamefont {P.}~\bibnamefont {Czarnik}},\ and\
  \bibinfo {author} {\bibfnamefont {L.}~\bibnamefont {Cincio}},\ }\bibfield
  {title} {\bibinfo {title} {{Precise Extrapolation of the Correlation Function
  Asymptotics in Uniform Tensor Network States with Application to the
  Bose-Hubbard and XXZ Models}},\ }\href
  {https://doi.org/10.1103/PhysRevX.8.041033} {\bibfield  {journal} {\bibinfo
  {journal} {Phys. Rev. X}\ }\textbf {\bibinfo {volume} {8}},\ \bibinfo {pages}
  {041033} (\bibinfo {year} {2018})}\BibitemShut {NoStop}%
\bibitem [{\citenamefont {Vanhecke}\ \emph {et~al.}(2019)\citenamefont
  {Vanhecke}, \citenamefont {Haegeman}, \citenamefont {Van~Acoleyen},
  \citenamefont {Vanderstraeten},\ and\ \citenamefont
  {Verstraete}}]{vanhecke_scaling_hypothesis_2019}%
  \BibitemOpen
  \bibfield  {author} {\bibinfo {author} {\bibfnamefont {B.}~\bibnamefont
  {Vanhecke}}, \bibinfo {author} {\bibfnamefont {J.}~\bibnamefont {Haegeman}},
  \bibinfo {author} {\bibfnamefont {K.}~\bibnamefont {Van~Acoleyen}}, \bibinfo
  {author} {\bibfnamefont {L.}~\bibnamefont {Vanderstraeten}},\ and\ \bibinfo
  {author} {\bibfnamefont {F.}~\bibnamefont {Verstraete}},\ }\bibfield  {title}
  {\bibinfo {title} {{Scaling Hypothesis for Matrix Product States}},\ }\href
  {https://doi.org/10.1103/PhysRevLett.123.250604} {\bibfield  {journal}
  {\bibinfo  {journal} {Phys. Rev. Lett.}\ }\textbf {\bibinfo {volume} {123}},\
  \bibinfo {pages} {250604} (\bibinfo {year} {2019})}\BibitemShut {NoStop}%
\bibitem [{\citenamefont {Zauner}\ \emph {et~al.}(2015)\citenamefont {Zauner},
  \citenamefont {Draxler}, \citenamefont {Vanderstraeten}, \citenamefont
  {Degroote}, \citenamefont {Haegeman}, \citenamefont {Rams}, \citenamefont
  {Stojevic}, \citenamefont {Schuch},\ and\ \citenamefont
  {Verstraete}}]{zauner_transfer_2015}%
  \BibitemOpen
  \bibfield  {author} {\bibinfo {author} {\bibfnamefont {V.}~\bibnamefont
  {Zauner}}, \bibinfo {author} {\bibfnamefont {D.}~\bibnamefont {Draxler}},
  \bibinfo {author} {\bibfnamefont {L.}~\bibnamefont {Vanderstraeten}},
  \bibinfo {author} {\bibfnamefont {M.}~\bibnamefont {Degroote}}, \bibinfo
  {author} {\bibfnamefont {J.}~\bibnamefont {Haegeman}}, \bibinfo {author}
  {\bibfnamefont {M.~M.}\ \bibnamefont {Rams}}, \bibinfo {author}
  {\bibfnamefont {V.}~\bibnamefont {Stojevic}}, \bibinfo {author}
  {\bibfnamefont {N.}~\bibnamefont {Schuch}},\ and\ \bibinfo {author}
  {\bibfnamefont {F.}~\bibnamefont {Verstraete}},\ }\bibfield  {title}
  {\bibinfo {title} {{Transfer matrices and excitations with matrix product
  states}},\ }\href {https://doi.org/10.1088/1367-2630/17/5/053002} {\bibfield
  {journal} {\bibinfo  {journal} {New J. Phys.}\ }\textbf {\bibinfo {volume}
  {17}},\ \bibinfo {pages} {053002} (\bibinfo {year} {2015})}\BibitemShut
  {NoStop}%
\bibitem [{\citenamefont {He}\ \emph {et~al.}(2017)\citenamefont {He},
  \citenamefont {Zaletel}, \citenamefont {Oshikawa},\ and\ \citenamefont
  {Pollmann}}]{he_signatures_2017}%
  \BibitemOpen
  \bibfield  {author} {\bibinfo {author} {\bibfnamefont {Y.-C.}\ \bibnamefont
  {He}}, \bibinfo {author} {\bibfnamefont {M.~P.}\ \bibnamefont {Zaletel}},
  \bibinfo {author} {\bibfnamefont {M.}~\bibnamefont {Oshikawa}},\ and\
  \bibinfo {author} {\bibfnamefont {F.}~\bibnamefont {Pollmann}},\ }\bibfield
  {title} {\bibinfo {title} {Signatures of {{Dirac Cones}} in a {{DMRG Study}}
  of the {{Kagome Heisenberg Model}}},\ }\href
  {https://doi.org/10.1103/PhysRevX.7.031020} {\bibfield  {journal} {\bibinfo
  {journal} {Physical Review X}\ }\textbf {\bibinfo {volume} {7}},\ \bibinfo
  {pages} {031020} (\bibinfo {year} {2017})}\BibitemShut {NoStop}%
\bibitem [{\citenamefont {Hastings}(2004)}]{hastings_locality_2004}%
  \BibitemOpen
  \bibfield  {author} {\bibinfo {author} {\bibfnamefont {M.~B.}\ \bibnamefont
  {Hastings}},\ }\bibfield  {title} {\bibinfo {title} {Locality in {Quantum}
  and {Markov} {Dynamics} on {Lattices} and {Networks}},\ }\href
  {https://doi.org/10.1103/PhysRevLett.93.140402} {\bibfield  {journal}
  {\bibinfo  {journal} {Phys. Rev. Lett.}\ }\textbf {\bibinfo {volume} {93}},\
  \bibinfo {pages} {140402} (\bibinfo {year} {2004})}\BibitemShut {NoStop}%
\bibitem [{\citenamefont {Bethe}(1931)}]{bethe_1931}%
  \BibitemOpen
  \bibfield  {author} {\bibinfo {author} {\bibfnamefont {H.}~\bibnamefont
  {Bethe}},\ }\bibfield  {title} {\bibinfo {title} {{Zur Theorie der
  Metalle}},\ }\href {https://doi.org/10.1007/BF01341708} {\bibfield  {journal}
  {\bibinfo  {journal} {Z. Phys.}\ }\textbf {\bibinfo {volume} {71}},\ \bibinfo
  {pages} {205} (\bibinfo {year} {1931})}\BibitemShut {NoStop}%
\bibitem [{\citenamefont {Yamada}(1969)}]{yamada_fermi-liquid_1969}%
  \BibitemOpen
  \bibfield  {author} {\bibinfo {author} {\bibfnamefont {T.}~\bibnamefont
  {Yamada}},\ }\bibfield  {title} {\bibinfo {title} {{Fermi-Liquid Theory of
  Linear Antiferromagnetic Chains}},\ }\href
  {https://doi.org/10.1143/PTP.41.880} {\bibfield  {journal} {\bibinfo
  {journal} {Progress of Theoretical Physics}\ }\textbf {\bibinfo {volume}
  {41}},\ \bibinfo {pages} {880} (\bibinfo {year} {1969})}\BibitemShut
  {NoStop}%
\bibitem [{\citenamefont {M\"uller}\ \emph {et~al.}(1981)\citenamefont
  {M\"uller}, \citenamefont {Thomas}, \citenamefont {Beck},\ and\ \citenamefont
  {Bonner}}]{mueller_quantum_spin_dynamics_1981}%
  \BibitemOpen
  \bibfield  {author} {\bibinfo {author} {\bibfnamefont {G.}~\bibnamefont
  {M\"uller}}, \bibinfo {author} {\bibfnamefont {H.}~\bibnamefont {Thomas}},
  \bibinfo {author} {\bibfnamefont {H.}~\bibnamefont {Beck}},\ and\ \bibinfo
  {author} {\bibfnamefont {J.~C.}\ \bibnamefont {Bonner}},\ }\bibfield  {title}
  {\bibinfo {title} {Quantum spin dynamics of the antiferromagnetic linear
  chain in zero and nonzero magnetic field},\ }\href
  {https://doi.org/10.1103/PhysRevB.24.1429} {\bibfield  {journal} {\bibinfo
  {journal} {Phys. Rev. B}\ }\textbf {\bibinfo {volume} {24}},\ \bibinfo
  {pages} {1429} (\bibinfo {year} {1981})}\BibitemShut {NoStop}%
\bibitem [{\citenamefont {Karbach}\ \emph {et~al.}(1997)\citenamefont
  {Karbach}, \citenamefont {M\"uller}, \citenamefont {Bougourzi}, \citenamefont
  {Fledderjohann},\ and\ \citenamefont {M\"utter}}]{karbach_two-spinon_1997}%
  \BibitemOpen
  \bibfield  {author} {\bibinfo {author} {\bibfnamefont {M.}~\bibnamefont
  {Karbach}}, \bibinfo {author} {\bibfnamefont {G.}~\bibnamefont {M\"uller}},
  \bibinfo {author} {\bibfnamefont {A.~H.}\ \bibnamefont {Bougourzi}}, \bibinfo
  {author} {\bibfnamefont {A.}~\bibnamefont {Fledderjohann}},\ and\ \bibinfo
  {author} {\bibfnamefont {K.-H.}\ \bibnamefont {M\"utter}},\ }\bibfield
  {title} {\bibinfo {title} {{Two-spinon dynamic structure factor of the
  one-dimensional $S=\frac{1}{2}$ Heisenberg antiferromagnet}},\ }\href
  {https://doi.org/10.1103/PhysRevB.55.12510} {\bibfield  {journal} {\bibinfo
  {journal} {Phys. Rev. B}\ }\textbf {\bibinfo {volume} {55}},\ \bibinfo
  {pages} {12510} (\bibinfo {year} {1997})}\BibitemShut {NoStop}%
\bibitem [{Note1()}]{Note1}%
  \BibitemOpen
  \bibinfo {note} {Apart from an intrinsic broadening set by the longest times
  reached within the numerical simulations.}\BibitemShut {Stop}%
\bibitem [{\citenamefont {Knolle}\ \emph {et~al.}(2014)\citenamefont {Knolle},
  \citenamefont {Kovrizhin}, \citenamefont {Chalker},\ and\ \citenamefont
  {Moessner}}]{knolle_dynamics_2014}%
  \BibitemOpen
  \bibfield  {author} {\bibinfo {author} {\bibfnamefont {J.}~\bibnamefont
  {Knolle}}, \bibinfo {author} {\bibfnamefont {D.}~\bibnamefont {Kovrizhin}},
  \bibinfo {author} {\bibfnamefont {J.}~\bibnamefont {Chalker}},\ and\ \bibinfo
  {author} {\bibfnamefont {R.}~\bibnamefont {Moessner}},\ }\bibfield  {title}
  {\bibinfo {title} {Dynamics of a {Two}-{Dimensional} {Quantum} {Spin}
  {Liquid}: {Signatures} of {Emergent} {Majorana} {Fermions} and {Fluxes}},\
  }\href {https://doi.org/10.1103/PhysRevLett.112.207203} {\bibfield  {journal}
  {\bibinfo  {journal} {Phys. Rev. Lett.}\ }\textbf {\bibinfo {volume} {112}},\
  \bibinfo {pages} {207203} (\bibinfo {year} {2014})}\BibitemShut {NoStop}%
\bibitem [{\citenamefont {Knolle}\ \emph {et~al.}(2015)\citenamefont {Knolle},
  \citenamefont {Kovrizhin}, \citenamefont {Chalker},\ and\ \citenamefont
  {Moessner}}]{knolle_dynamics_2015}%
  \BibitemOpen
  \bibfield  {author} {\bibinfo {author} {\bibfnamefont {J.}~\bibnamefont
  {Knolle}}, \bibinfo {author} {\bibfnamefont {D.~L.}\ \bibnamefont
  {Kovrizhin}}, \bibinfo {author} {\bibfnamefont {J.~T.}\ \bibnamefont
  {Chalker}},\ and\ \bibinfo {author} {\bibfnamefont {R.}~\bibnamefont
  {Moessner}},\ }\bibfield  {title} {\bibinfo {title} {Dynamics of
  fractionalization in quantum spin liquids},\ }\href
  {https://doi.org/10.1103/PhysRevB.92.115127} {\bibfield  {journal} {\bibinfo
  {journal} {Phys. Rev. B}\ }\textbf {\bibinfo {volume} {92}},\ \bibinfo
  {pages} {115127} (\bibinfo {year} {2015})}\BibitemShut {NoStop}%
\end{thebibliography}%

\end{document}